\newcount\drawps\drawps=1 
\newcount\pssize
\ifnum\drawps=1
   \documentstyle[aps,twocolumn,prb,psfig]{revtex}
\else
   \documentstyle[aps,preprint,prb,psfig]{revtex}
\fi

\def\ba{ {\bf a} }  
\def\bn{ {\bf n} }
\def\bm{ {\bf m} }

\def\bk{ {\bf k} }  
\def\bq{ {\bf q} }
\def\bp{ {\bf p} }

\begin{document}
\draft

\title{Spinwave damping in the two-dimensional ferromagnetic XY model}
\author{G.\ M.\  Wysin,\cite{address1} 
M.\ E.\  Gouv\^ea, and A.\ S.\ T.\ Pires}
\address{
Departamento de F\'{\i}sica, ICEx,
Universidade Federal de Minas Gerais
Belo Horizonte, \\CP 702, CEP 30123-970, MG, Brazil
}
\date{December 6, 1999}
\maketitle

\begin{abstract} 
The effect of damping of spinwaves in a two-dimensional 
classical ferromagnetic XY model is considered.   
The damping rate $\Gamma_{\bq}$ is calculated using the leading 
diagrams due to the quartic-order deviations from the harmonic spin 
Hamiltonian.   
The resulting four-dimensional integrals are evaluated by extending the 
techniques developed by Gilat and others for spectral density types of 
integrals.   
$\Gamma_{\bq}$ is included into the memory function formalism due to Reiter 
and Solander, and Menezes, to determine the dynamic structure function 
$S(\bq,\omega)$.  
For the infinite sized system, the memory function approach is found to give 
non-divergent spinwave peaks, and a smooth nonzero background intensity 
(``plateau'' or distributed intensity) for the whole range of frequencies 
below the spinwave peak. 
The background amplitude relative to the spinwave peak rises with temperature,
and eventually becomes higher than the spinwave peak, where it appears as a 
central peak.  
For finite-sized systems,  there are multiple sequences of weak peaks on 
both sides of the spinwave peaks whose number and positions depend on the 
system size and wavevector in integer units of $2\pi/L$.
These dynamical finite size effects are explained in the memory function
analysis as due to either spinwave difference processes below the spinwave
peak or sum processes above the spinwave peak.
These features are also found in classical Monte Carlo -- Spin-Dynamics 
simulations.    
\end{abstract} 
\pacs{PACS numbers: 75.10.Hk,75.30.Ds,75.40.Gb,75.40.Mg}

\section{Introduction}
The two-dimensional (2D) ferromagnetic XY model has had a long history
of studies concerning the dynamics of spinwaves at low temperatures. 
It was first analyzed by Villain\cite{Villain74} and by
Moussa and Villain\cite{Moussa+76}, for low temperatures,
in the harmonic approximation.
They found that the dynamic structure function $S(\bq,\omega)$
has a spinwave peak of the form
\begin{equation}
S(\bq,\omega) \sim \vert \omega - \omega_{\bq} \vert^{-1+\eta/2} ~ ,
\end{equation}
where $\eta$ is the critical exponent describing the decay of the
static spin-spin correlation function, and $\omega_{\bq}$ is the spinwave
frequency at wavevector $\bq$.
Nelson and Fisher\cite{Nelson+77} treated the model in a fixed-length
hydrodynamic description,  obtaining the following expression around the 
spinwave peak,
\begin{equation}
S(\bq,\omega)  \sim \frac{1}{q^{3-\eta}} 
\frac{1}{\vert 1-\omega^{2} / \omega_{\bq}^{2} \vert^{1-\eta}} ~ ,
\end{equation}
where for small $q$, $\omega_{\bq}=cq$, with $c$ being the spinwave
velocity.
These theoretical approaches led to spinwave peaks diverging 
unphysically at the spinwave frequency.
Here we describe an alternative theoretical description in
which the divergence does not occur.

At higher temperatures, the spinwave or harmonic approximation
begins to fail,  and XY magnets are known to undergo a
topological phase transition\cite{Berezinskii70,Kosterlitz73}
 associated with the unbinding of vortices.
We generally consider temperatures below this unbinding temperature
$T_{\rm BKT}$, where a spinwave approximation will be valid.
At very low temperatures any nonlinear effects should be very weak,
do not lead to vortex production, and can be included as the
damping or scattering of the spinwave modes, that results in
a spinwave lifetime. 
Thus our approach will be to start from the harmonic approximation
to the spin Hamiltonian, $H_0$, and then include the damping that results 
from the terms fourth order in spin operators  ($H_1$) in an expansion of
the original full spin Hamiltonian, $H$.
The resulting damping strength $\Gamma_{\bq}$ will imply a finite lifetime 
$\tau=\Gamma_{\bq}^{-1}$ for the spinwave modes.  
We are interested to investigate whether the presence of such
lifetime effects can eliminate the divergence of the dynamic structure 
function $S(\bq,\omega)$ at the spinwave frequency.

We calculate the spinwave damping by a Feynmann diagram technique for 
finite temperatures, applied to the fourth-order Hamiltonian $H_1$, using 
the leading diagrams to second order in $H_1$.  
The only diagrams we include are those that give a contribution to the 
{\em imaginary part} of the spinwave energy, i.e., to the damping or inverse 
lifetime.
These are the terms that involve a {\em dynamical} effect on the
spinwave energies.
The associated integral expressions for these diagrams require
the evaluation of four-dimensional integrals involving a delta-function.
Methods for evaluating this type of integral in three dimensions
have been developed by Gilat and co-workers; here we show how
to extend their work and apply the same technique in four dimensions.
%

In principle, we could also use the Feynmann technique to determine
the leading changes in the {\em real part} of the spinwave
energy, $\Delta \omega_{\bq}$, due to the perturbation $H_1$.  
This real part would result as the usual effect due to finite
temperature:  a shift in the spinwave frequencies to lower
values as the temperature increases.
It is essentially a {\em static} effect, primarily due to the
renormalization of the nearest neighbor exchange interaction strength $J$
with temperature.
Calculation of these spinwave frequency shifts by the Feynmann technique 
would require including diagrams first order in $H_1$; the so-called
single-loop diagrams.
In this paper, however, we are primarily concerned with accounting for 
dynamical effects.  
We will include the renormalization of the spinwave energies
using the memory function formalism, combined with some input 
information from classical Monte Carlo (MC) data.   
In the memory function approach,  the position of the spinwave peak
in $S(\bq,\omega)$ does not occur exactly at the zero-temperature
dispersion relation, $\omega_{\bq}$, but rather, close to the value
$\sqrt{\langle \omega^2 \rangle_{\bq}} < \omega_{\bq}$, where 
$\langle \omega^2 \rangle_{\bq}$ is the temperature-dependent second moment.
$\langle \omega^2 \rangle_{\bq}$ can be estimated from static quantities,
either using low-temperature analytic expressions or results from
Monte Carlo, the latter being necessary for temperatures 
approaching $T_{\rm BKT}$.
The memory function approach already includes to leading order 
the exchange strength renormalization, as can be seen in the shift 
in the position of the spinwave peaks. 
Thus we use it rather than making any renormalization of the
exchange $J$.

A few comments can be made about how the damping $\Gamma_{\bq}$ is 
included into the memory function calculation.
Note that the damping calculation sketched above is
completely independent of the memory function calculation;
the latter is a formalism for calculating dynamic correlations
for a given Hamiltonian and operators.  
Due to the XY symmetry of the model, the natural coordinates 
of the spins are the in-plane angle $\phi$ and out-of-plane 
spin component $S^z$.  
Fourier transforming and using a standard canonical transformation,
the Hamiltonian is written in terms of creation and 
annihilation operators $a_{\bq}^{\dagger}$ and $a_{\bq}$, with time dependencies
varying as $e^{i\omega_{\bq} t}$ and $e^{-i\omega_{\bq} t}$, respectively.
Once damping is included, the time dependencies become
$e^{i(\omega_{\bq}+i\Gamma_{\bq})t}$ and $e^{-i(\omega_{\bq}-i\Gamma_{\bq})t}$, 
respectively.
Henceforth the calculation proceeds as usual in the memory function 
approach, except for this small modification of the operator
time dependencies via the shifted operator frequencies,
$\omega_{\bq}\rightarrow \omega_{\bq} \pm i \Gamma_{\bq}$.
%

In the memory function calculation, spinwave sum and difference
processes enter.
These are characterized by combinations of frequencies,
$\Omega_{\pm}=\omega_{\frac{\bq}{2}+{\bk}}\pm 
\omega_{\frac{\bq}{2}-{\bk}}$,
where $\bq$ is the wavevector at which the dynamic correlations
are measured, and $\bk$ is summed over to include all possible
such processes.
Based on the above modifications of the operator frequencies,
there also results the modification for these processes,
$\Omega_{\pm}\rightarrow \Omega_{\pm}+i(\Gamma_{\frac{\bq}{2}+\bk}
+ \Gamma_{\frac{\bq}{2}-\bk} )$.
The damping rates for different wavevectors always add for 
both types of processes.
%

The paper is organized as follows:
First, in Sec.\ \ref{Model}, we present the model and develop 
the expansion of the Hamiltonian to quartic order in 
creation/annihilation operators.
In Sec.\ \ref{Diagrams}, the expressions for the damping 
via the Feynmann technique are presented.
Then we show how these are evaluated using the Gilat procedure.
In Sec.\ \ref{Memory}, the key aspects of the memory function calculation
of $S(\bq,\omega)$ are reviewed, together with a focus on the modification 
due to the damping. 
Some details of the Monte Carlo and spin dynamics simulations
are summarized in Sec.\ \ref{MC+SD}.
This is followed in Sec.\ \ref{Results} by presentation of the results 
both for very large and relatively smaller systems, in order to discuss 
the dynamical finite size effects.
The memory function results also are compared with standard 
Monte Carlo -- spin-dynamics calculations of the dynamic correlation
functions in order to evaluate the validity and accuracy of the 
method. 
Our conclusions are presented in Sec.\ \ref{Conclusions}.

\section{The Model}
\label{Model}
The ferromagnetic spin Hamiltonian under consideration is
\begin{equation}
\label{Hamil}
H = -\frac{J}{2} \sum_{\bn} \sum_{\ba} 
\left( S^x_{\bn} S^x_{\bn+\ba} + S^y_{\bn} S^y_{\bn+\ba} 
+ \lambda S^z_{\bn} S^z_{\bn+\ba}  \right) ~ , 
\end{equation}
where the first sum is over all sites $\bn$ of a 2D square lattice, 
the second is over all nearest neighbor displacements $\ba$,
$J$ is the nearest neighbor coupling strength, and each ${\bf S}_{\bn}$ 
is a three-component classical spin-vector.
Because each bond is counted twice, a factor of $\frac{1}{2}$ is
included on $J$.
We primarily concentrate on the XY model, with anisotropy
parameter $\lambda=0$, however, for completeness, formulas
covering the easy-plane anisotropy range, $0\le \lambda < 1$,
are presented.
%

\subsection{Expansion of Hamiltonian}
In order to develop the damping as a perturbation of the harmonic
approximation, we need to expand the full spin Hamiltonian, 
including the harmonic (quadratic) and quartic deviations from 
the ground state.
The ground state corresponds to spins aligned in any direction within 
the XY plane.
Writing the spins in Hamiltonian (\ref{Hamil}) using coordinates
$\phi_{\bn}$ and $z_{\bn}$, 
\begin{equation}
\label{mapping}
{\bf S}_{\bn} = S(\sqrt{1-z^2_{\bn}} \cos\phi_{\bn}, 
\sqrt{1-z^2_{\bn}} \sin\phi_{\bn}, z_{\bn} ) ~ ,
\end{equation}
the Hamiltonian becomes
\begin{equation}
H=-\frac{JS^2}{2} \sum_{\bn,\bm} \left[ 
\sqrt{(1-z^2_{\bn})(1-z^2_{\bm})} \cos(\phi_{\bn}-\phi_{\bm})
+ \lambda z_{\bn} z_{\bm} \right] ~ ,
\end{equation}
where we use a simplified notation for the nearest neighbors, 
${\bm}={\bn}+{\ba}$.
At low temperature, it is reasonable to expand for small $z_{\bn}$ and 
$(\phi_{\bn}-\phi_{\bm})$, keeping up to quartic order in these, obtaining,
\begin{equation}
H = H_0+H_1
\end{equation}
where the harmonic and quartic Hamiltonians are
\begin{eqnarray}
H_0 &=& -\frac{JS^2}{2} \sum_{\bn,\bm} 
\Bigl[ 1+\lambda z_{\bn} z_{\bm}  \nonumber \\
&& -\frac{1}{2}(z^2_{\bn}+z^2_{\bm}) 
-\frac{1}{2} (\phi_{\bn}-\phi_{\bm})^2 \Bigr],
\end{eqnarray}
\begin{eqnarray}
H_1 &=& \frac{JS^2}{2} \sum_{\bn,\bm} \Bigl[ 
 \frac{1}{8} (z^4_{\bn}+z^4_{\bm}-2z^2_{\bn} z^2_{\bm} )  \nonumber \\
&& -\frac{1}{4} (z^2_{\bn} + z^2_{\bm})(\phi_{\bn} - \phi_{\bm})^2 
-\frac{1}{4!}(\phi_{\bn} - \phi_{\bm})^4
\Bigr] ~ .
\end{eqnarray}

It is convenient to carry out the calculations in
wavevector space by defining Fourier transforms
\begin{equation}
\label{Fourier}
\phi_{\bn} = \frac{1}{\sqrt{N}} \sum_{\bq} 
e^{i \bq \cdot \bn} \phi_{\bq} ~  ,
\end{equation}
together with a similar definition for $z_{\bq}$. 
Here $\bq$ denotes a two-dimensional wavevector, 
the sums are over the first Brillouin zone (BZ),
and $N$ is the total number of lattice sites.
Then we rewrite $H_0$ and $H_1$ as
\begin{equation}
H_0 = 2JS^2 \sum_{\bq} \left[ (1-\gamma_{\bq}) \phi_{\bq} \phi_{-\bq} 
+ (1-\lambda \gamma_{\bq}) z_{\bq} z_{-\bq} \right] ~ ,
\end{equation}
and
\begin{eqnarray}
\label{H1}
H_1 & = & \frac{JS^2}{N}  
\sum_{\bq_1,\bq_2,\bq_3,\bq_4} ~ \Bigl\{
\frac{1}{2}(1-\gamma_{\bq_1+\bq_2}) 
z_{\bq_1} z_{\bq_2} z_{\bq_3} z_{\bq_4} \nonumber \\
 & - &
(1+\gamma_{\bq_3+\bq_4} -\gamma_{\bq_3} -\gamma_{\bq_4}) 
z_{\bq_1}z_{\bq_2}\phi_{\bq_3}\phi_{\bq_4}
\nonumber \\
 & - & \frac{1}{6}
(1+\gamma_{\bq_1+\bq_2}+\gamma_{\bq_1+\bq_3}
+ \gamma_{\bq_1+\bq_4} \nonumber \\
& - & \gamma_{\bq_1}-\gamma_{\bq_2}-\gamma_{\bq_3}
-\gamma_{\bq_4})
\phi_{\bq_1}\phi_{\bq_2}\phi_{\bq_3}\phi_{\bq_4} \Bigr\} ~ .
\end{eqnarray}
For the square lattice, with coordination number $z=4$,
the $\gamma_{\bq}$ function
\begin{equation}
\gamma_{\bq} = \frac{1}{z}\sum_{\ba} e^{i \bq \cdot \ba} 
= \frac{1}{2} (\cos q_x + \cos q_y )
\end{equation}
is an even function of $\bq$, i.e., $\gamma_{\bq}=\gamma_{-\bq}$.
The sums in (\ref{H1}) are constrained to 
$\bq_1+\bq_2+\bq_3+\bq_4=0$.
%

Using the following canonical transformation to 
creation/annihilation operators,
\begin{equation}
\label{canonical}
\phi_{\bq}=\alpha_{\bq} (a^{\dagger}_{\bq} + a_{-\bq}), \quad\quad
z_{\bq}= i\beta_{\bq} (a^{\dagger}_{\bq} - a_{-\bq}) ~ ,
\end{equation}
\begin{equation}
\alpha_{\bq}=\left[\frac{1-\lambda \gamma_{\bq}}
{4S^2(1-\gamma_{\bq})}\right]^{1/4}, \quad\quad
\beta_{\bq}=\left[\frac{1-\gamma_{\bq}}
{4S^2(1-\lambda\gamma_{\bq})}\right]^{1/4} ~ ,
\end{equation}
the quadratic Hamiltonian is diagonalized,
\begin{equation}
H_0 =  \sum_{\bq} \omega_{\bq} 
\left[ a^{\dagger}_{\bq} a_{\bq} + \frac{1}{2} \right] ~ ,
\end{equation}
\begin{equation} 
\omega_{\bq} = 4JS \sqrt{(1-\gamma_{\bq})(1-\lambda\gamma_{\bq})} ~ .
\end{equation}
The diagonal parts of $H_1$ can then be written as
\begin{equation}
\label{H_1}
H_1 =  \sum_{\bq_1+\bq_2+\bq_3+\bq_4=0} 
A(\bq_1,\bq_2,\bq_3,\bq_4) 
~~ a^{\dagger}_{\bq_1} a^{\dagger}_{\bq_2} a_{-\bq_3} a_{-\bq_4},
\end{equation}
where
\begin{eqnarray}
&A&(\bq_1,\bq_2,\bq_3,\bq_4) =  
\frac{JS^2}{N} \Bigl\{  \nonumber \\
&& \beta_{\bq_1}\beta_{\bq_2}\beta_{\bq_3}\beta_{\bq_4}
(3-\gamma_{\bq_1+\bq_2}-\gamma_{\bq_1+\bq_3}
-\gamma_{\bq_1+\bq_4}) \nonumber \\
&-& 3\beta_{\bq_1}\alpha_{\bq_2}\alpha_{\bq_3}\beta_{\bq_4} 
(1+\gamma_{\bq_2+\bq_3}-\gamma_{\bq_2}-\gamma_{\bq_3}) \nonumber \\
&-& \beta_{\bq_2}\alpha_{\bq_1}\alpha_{\bq_4}\beta_{\bq_3}
(1+\gamma_{\bq_1+\bq_4}-\gamma_{\bq_1}
-\gamma_{\bq_4}) \nonumber \\
&+& \beta_{\bq_3}\alpha_{\bq_1}\alpha_{\bq_2}\beta_{\bq_4}
(1+\gamma_{\bq_1+\bq_2}-\gamma_{\bq_1}-\gamma_{\bq_2}) \nonumber \\
&+& \beta_{\bq_1}\alpha_{\bq_3}\alpha_{\bq_4}\beta_{\bq_2}
(1+\gamma_{\bq_3+\bq_4}-\gamma_{\bq_3}-\gamma_{\bq_4}) \nonumber \\
&-& \alpha_{\bq_1}\alpha_{\bq_2}\alpha_{\bq_3}\alpha_{\bq_4}
(1+\gamma_{\bq_1+\bq_2}+\gamma_{\bq_1+\bq_3}
+\gamma_{\bq_1+\bq_4} \nonumber \\
&-&\gamma_{\bq_1} -\gamma_{\bq_2}-\gamma_{\bq_3}-\gamma_{\bq_4})  
\Bigr\} ~ .
\end{eqnarray}
We have dropped terms in $H_1$ whose expectation values in 
the ground state are zero.
Note also that if a particular $\bq_i$ is zero, 
then $\alpha_{\bq_i}$ diverges, however, when this occurs the 
corresponding factors multiplying $\alpha_{\bq_i}$ 
go to zero faster, canceling the divergence.
%

\section{Spinwave Damping due to $H_1$}
\label{Diagrams}
The time evolution of the creation/annihilation operators as
given from the harmonic Hamiltonian is
\begin{equation}
a_{\bq}^{\dagger}(t) = a_{\bq}^{\dagger}(0) e^{i\omega_{\bq} t}, \quad\quad
a_{\bq}(t) = a_{\bq}(0) e^{-i\omega_{\bq} t} ~ . 
\end{equation}
As discussed in the Introduction, the effect of the $H_1$ perturbation
is to shift the spinwave frequencies, in the most general case, by some 
complex self-energy $\Sigma_{\bq}$.
For the moment, we ignore the real part of $\Sigma_{\bq}$, whose effect will 
be included by the memory function approach, and concentrate instead on the 
contributions to the imaginary part of $\Sigma_{\bq}$, which determine the 
spinwave damping.
To find $\Sigma_{\bq}$, the contributions of $H_1$ to the 
single particle Green's function will be evaluated, keeping only 
the diagrams that have a nonzero imaginary part. 
The lowest order diagrams that satisfy this requirement are
found to be second order in $H_1$; there is no contribution 
from first order diagrams.

Following the standard procedure,\cite{Mattuck,Fetter,Abrikosov} 
the time-ordered zero-temperature Green's function, in second order, is
\begin{eqnarray}
&i&G_2(\bk_2,\bk_1,t_2-t_1) = 
\frac{(-i)^2}{2!}  \int_{-\infty}^{\infty} dt_1^{\prime}
\int_{-\infty}^{\infty} dt_2^{\prime} \times \nonumber \\ 
&&\langle \Phi_0 \vert T \{ H_1(t_1^{\prime}) H_1(t_2^{\prime}) 
a_{\bk_2}(t_2) a_{\bk_1}^{\dagger}(t_1) \} \vert \Phi_0 \rangle ~ .
\end{eqnarray}
Here $T$ is the time ordering operator, and $\Phi_0$ is the ground state
of $H_0$.
Physically, this Green function represents the amplitude for a quasiparticle
(i.e., spinwave) created at time $t_1$ in state $\bk_1$ to be destroyed 
at time $t_2$ in state $\bk_2$, after undergoing two scattering events
due to $H_1$.
The free-particle Green function,
\begin{equation}
i G_0(\bk_2,\bk_1,t_2-t_1)=
\langle \Phi_0 \vert T \{ 
a_{\bk_2}(t_2) a_{\bk_1}^{\dagger}(t_1) \} \vert \Phi_0 \rangle ~ ,
\end{equation}
involves no scattering. 
Due to momentum conservation, these Green functions are proportional 
to the Kronecker delta, $\delta_{\bk_1,\bk_2}$, so we can drop 
the double momentum indices.
Also, because of time translational invariance, they depend only on the
time difference $t=t_2-t_1$.
Introducing the time Fourier transforms, 
\begin{equation} 
G_n(\bk,t)=\frac{1}{2\pi} \int_{-\infty}^{\infty} 
d\omega ~ G_n(\bk,\omega) e^{-i \omega t} ~ ,
\end{equation}
the free propagator becomes the well-known expression,
\begin{equation}
G_0(\bk,\omega)= \frac{1}{\omega-\omega_{\bk}+i\delta},
\end{equation}
where $\delta=0^{+}$.

\begin{figure}
\hskip 1.6in
\psfig{figure=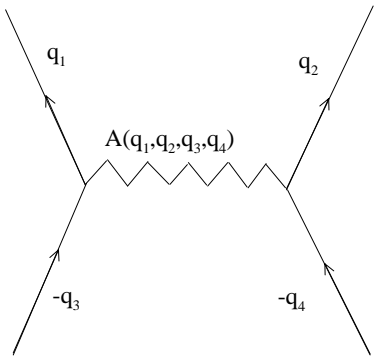,angle=-90.0,width=\pssize pt}
\caption{
\label{Aq1q2q3q4}
Notation for the representation of the interaction terms in $H_1$,
Eq.\ (\protect\ref{H_1}), in the Feynmann diagrams. 
}
\end{figure}

\begin{figure}
\hskip 1.6in
\psfig{figure=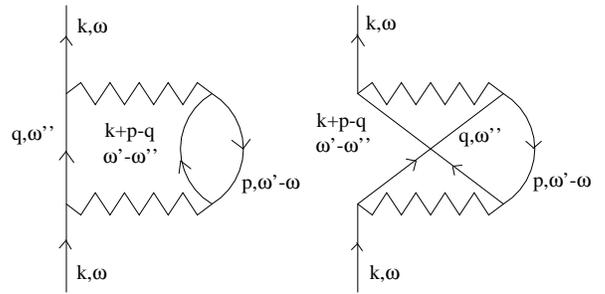,angle=-90.0,width=\pssize pt}
\caption{
\label{diagrams}
The two leading diagrams that contribute to $G_2(\bk,\omega)$,
as in Eqs. (\protect\ref{G_2}) and (\protect\ref{Sigma}).
}
\end{figure}

We apply the rules for Feynmann diagrams to evaluate the
full propagator, $G(\bk,\omega)$, in terms of $G_0$ and $G_2$. 
The free propagator $G_0(\bk,\omega)$ is represented as a straight line
with an associated momentum $\bk$ and frequency $\omega$.
The interaction term in (\ref{H_1}) is represented
as a wavy line as in Fig.\ \ref{Aq1q2q3q4}, with an outgoing straight line
(free propagator) for each creation operator and an incoming straight line 
for each destruction operator, both with assigned momentum and frequency.
The two leading diagrams that contribute to $G_2(\bk,\omega)$ 
are shown in Fig.\ \ref{diagrams}. 
Following the standard Feynmann rules for these diagrams, with
summations over the intermediate frequencies and wavevectors,  
and using momentum and frequency conservation at each junction,
$G_2(\bk,\omega)$ can be written as
\begin{equation}
\label{G_2}
G_2(\bk,\omega)= 
G_0(\bk,\omega)\Sigma(\bk,\omega)G_0(\bk,\omega) ~ ,
\end{equation}
where the self-energy function due to the scattering parts of the
diagrams is
\begin{eqnarray}
\label{Sigma}
\Sigma&&(\bk,\omega) = -4 \sum_{\bp,\bq} 
\int \frac{d\omega^{\prime}}{2\pi} \int \frac{d\omega^{\prime\prime}}{2\pi} 
\nonumber \\
&&G_0(\bq,\omega^{\prime\prime}) ~  G_0(\bp,\omega^{\prime}-\omega) 
 ~ G_0(\bk+\bp-\bq,\omega^{\prime}-\omega^{\prime\prime}) \nonumber \\
&& A(\bq,\bk+\bp-\bq,-\bk,-\bp) \times \nonumber \\
&& \left[ A(\bk,\bp,-\bq,-\bk-\bp+\bq) 
+ A(\bp,\bk,-\bq,-\bk-\bp+\bq) \right] ~ .
\end{eqnarray}
$\Sigma(\bk,\omega)$ is a sum over the ``direct'' and the ``exchanged''
diagrams; the factor of 4 is the multiplicity of each of these under
rearranging the exterior lines.
$\Sigma(\bk,\omega)$ is seen to be $G_2(\bk,\omega)$ with its incoming and
outgoing lines removed.

The full interacting propagator $G(\bk,\omega)$ can be approximated as 
a sum over repeated applications of $\Sigma G_0$, i.e., 
\begin{equation}
G(\bk,\omega) = G_0+G_0(\Sigma G_0) + G_0(\Sigma G_0)^2 + ...,
\end{equation}
which, after summing, leads to the well-known Dyson equation,
\begin{equation}
\label{Dyson}
G(\bk,\omega) = \frac{1}{G_0^{-1}-\Sigma}
= \frac{1}{\omega-\omega_{\bk} +i\delta - \Sigma(\bk,\omega)} .
\end{equation}
The location of the pole gives the perturbed quasiparticle energy,
$\omega_{\bk}^{\prime}=\omega_{\bk}+{\rm Re}\Sigma(\bk,\omega_{\bk})$, 
and the damping $\Gamma_{\bk}=\Gamma(\bk,\omega_{\bk})$ is given by
\begin{equation}
\Gamma_{\bk} = - {\rm Im} ~ \Sigma(\bk,\omega_{\bk}) ~ .
\end{equation}
%

To apply the result to the system at finite temperature $T$,  and
obtain the self-energy associated with the temperature Green's function, 
the integrals over frequency are transformed to sums according to
\begin{equation}
\int \frac{d\omega^{\prime}}{2\pi} F(\omega^{\prime}) \rightarrow 
\frac{1}{\beta} \sum_{n=-\infty}^{\infty}F(i\omega_n) ~ , 
\end{equation}
\begin{equation}
\omega_n = \frac{2\pi}{\beta} n ~ , \quad\quad \beta=\frac{1}{k_B T}  ~ .
\end{equation}
Thus the frequency integrals in (\ref{Sigma}) become
\begin{eqnarray}
&& \int \frac{d\omega^{\prime}}{2\pi} 
\int \frac{d\omega^{\prime\prime}}{2\pi} ~  G_0 ~  G_0 ~ G_0 \nonumber \\
&& \rightarrow -
\frac{[n(\omega_q)-n(-\omega_{\bk+\bp-\bq})]
[n(\omega_{\bq}+\omega_{\bk+\bp-\bq})-n(\omega_{\bp})]}
{\omega-(\omega_{\bk+\bp-\bq}+\omega_{\bq}-\omega_{\bp})} ~ ,
\end{eqnarray}
where the Bose factors are
\begin{equation}
\label{Bose}
n(\omega) = (e^{\beta \omega}-1)^{-1} \approx 1/(\beta\omega) ~ .
\end{equation}
The latter form applies to the classical limit.
The imaginary part of $\Sigma$ is derived from the analytic relation,
\begin{equation}
\label{delta-relation}
\frac{1}{x+i\delta} = P\left(\frac{1}{x}\right)-i \pi \delta(x) ~ ,
\end{equation}
where $P$ takes the principal part and $\delta(x)$ is the
usual delta-function.
Because $\bq$ and $\bp$ each have two components, the sum in (\ref{Sigma})
is equivalent to a four-dimensional integration, once we take the 
continuum limit.
Each dimension converts as, for example, 
$\sum_{q_x} \rightarrow \frac{L_x}{2\pi} \int ~ dq_x$, where 
$L_x L_y = N a^2$ is the system size, in terms of the lattice 
constant $a$, leading to
\begin{equation}
\sum_{\bq} \sum_{\bp} \rightarrow \frac{N^2 a^2}{(2\pi)^4} 
\int_{-\pi}^{\pi} dq_x \int_{-\pi}^{\pi} dq_y 
\int_{-\pi}^{\pi} dp_x \int_{-\pi}^{\pi} dp_y ~ .
\end{equation}
The factors of $N$ here are canceled by those  in the expression
for $A(\bq_1,\bq_2,\bq_3,\bq_4)$, leading to a size-independent result.
Thus there finally results for the damping function,
\begin{equation}
\label{Gamma}
\Gamma(\bk,\omega) = \int d^2q \int d^2 p ~ F(\bq,\bp) ~ 
\delta(\omega-\omega_{\bk+\bp-\bq}+\omega_{\bp}-\omega_{\bq}) ~ ,
\end{equation}
where the kernel $F(\bq,\bp)$ is given by
\begin{eqnarray}
&&F(\bq,\bp) = 4\pi \left( \frac{L}{2\pi} \right)^4  
A(\bq,\bk+\bp-\bq,-\bk,-\bp) \times \nonumber \\
&&\left[ A(\bp,\bk,-\bq,-\bk-\bp+\bq) 
+ A(\bk,\bp,-\bq,-\bk-\bp+\bq) \right] \times \nonumber \\
&&[n(\omega_{\bq})-n(-\omega_{\bk+\bp-\bq})] 
[n(\omega_{\bq}+\omega_{\bk+\bp-\bq})-n(\omega_{\bp})] ~ .
\end{eqnarray}
We use this result to calculate the damping at the spinwave frequency: 
$\Gamma_{\bk}=\Gamma(\bk,\omega=\omega_{\bk})$, in the thermodynamic
limit $L\rightarrow\infty$.
Note that in the classical limit, the Bose factors are simply proportional
to temperature. 
Therefore, we only need to calculate $\Gamma_{\bk}$ at
one temperature ($T = 1.0 J$) and then scale with $T^2$ to get
its value at other temperatures.

\subsection{Evaluation of contour integrals in four dimensions}
\label{Integration}
The summation/integration in (\ref{Gamma}) involving a delta function presents
an interesting mathematical problem in its evaluation.
General integrals of this form (i.e., with one delta-function)
in $d$ dimensions can be considered,
\begin{equation}
\label{Integrals}
D(\omega) = \int d^d x ~ f({\bf x}) \delta(\omega-g({\bf x})) ,
\end{equation}
where ${\bf x}$ is a $d$-dimensional vector.
In two dimensions, for example, the delta-function will pick off
a particular contour that contributes to the integral; in three
dimensions that contour will actually be a surface, and in four
dimensions it will be a three-dimensional volume.   
Depending on the form of the function $g({\bf x})$, the contributing region 
need not be simply connected, but may consist of separated distinct parts.
For the higher dimensional cases, it is difficult to visualize the
region or regions that contribute to the integral. 
Thus it is necessary to have a way to evaluate the integral without
appealing directly to geometrical properties. 
%

In a series of papers, Gilat and co-workers\cite{Gilat66,Gilat69,Gilat72} 
have considered how to evaluate these integrals in three dimensions, 
in work related the calculation of spectral density functions.
We have used the ``linear analytic'' (LA) method\cite{Gilat69}
of Gilat and Raubenheimer, extended to four dimensions, which gives
good precision for few sample points, at little cost in complexity. 
Suppose we calculate $D(\omega)$ over a small range of $\omega$
and then take the limit $\omega\rightarrow\omega_{\bk}$.
To do so, a 4D cubic grid of small cells of size $(2b)^4$ is set 
in the ${\bf x}$-space.
Then we consider the contribution of each cell to $D(\omega)$,
thereafter summing over the cells.
Suppose the cells are small enough so that the kernel $f({\bf x})$
has only a small variation within a cell, and, therefore, we can use
its value at the cell center, ${\bf x}_c$.
Then we can write
\begin{equation}
\label{Gilat4d}
D(\omega)=\sum_{\rm cells} D_{\rm cell}(\omega) ~ ,
\end{equation}
\begin{equation}
\label{Gilatcell}
D_{\rm cell}(\omega)= 
f({\bf x}_c) \int_{\rm cell} d^4 x ~ \delta(\omega-g({\bf x})) ~ .
\end{equation}
The idea of the LA method is to expand $g({\bf x})$ around the cell center,
\begin{equation}
g({\bf x}) \approx g({\bf x}_c)
+({\bf x}-{\bf x}_c)\cdot \nabla g({\bf x}_c) ~ .
\end{equation}
Shifting coordinates to ${\bf u}={\bf x}-{\bf x}_c$, and writing 
the gradient at the cell center as 
$\nabla g({\bf x}_c)= \vert \nabla g({\bf x}_c) \vert \hat{\bf l}$,
where the unit vector $\hat{\bf l}=(l_1,l_2,l_3,l_4)$ points along 
the gradient evaluated at the cell center, we have
\begin{equation}
D_{\rm cell}(\omega)=\frac{f({\bf x}_c)}{\vert \nabla g({\bf x}_c) \vert}
\int_{\rm cell} d^4 u ~ \delta(w-{\bf u} \cdot \hat{\bf l}) ~ ,
\end{equation}
\begin{equation}
\label{w-omega}
w=\frac{\omega-g({\bf x}_c)}{\vert \nabla g({\bf x}_c) \vert} ~ .
\end{equation}
The variable $w$ is a rescaling of $\omega$. 
If we use the integral representation of the delta function,
in the form,
\begin{equation}
\delta(w-{\bf u} \cdot \hat{l}) = \frac{1}{2\pi} \int_{-\infty}^{\infty}
dk ~ e^{ik(w-u_1 l_1-u_2 l_2-u_3 l_3-u_4 l_4)}  ~ ,
\end{equation}
then the contribution from one cell is found to be
\begin{eqnarray}
\label{Dcell}
D_{\rm cell}(\omega) &=&  \frac{f({\bf x}_c)}{\vert \nabla g({\bf x}_c) \vert}
\frac{1}{2\pi} \int_{-\infty}^{\infty} dk ~ e^{ikw} 
\prod_{j=1}^{4} \int_{-b}^{b} du_j ~ e^{-ik u_j l_j} \nonumber \\
&=&  \frac{f({\bf x}_c)}{\vert \nabla g({\bf x}_c) \vert}
\frac{1}{12 \vert l_1 l_2 l_3 l_4 \vert} \sum_{i=1}^{16} (-1)^{m_i}
\vert w + w_i \vert^3 ~ ,
\end{eqnarray}
where the $w_i$ are the values of parameter $w$ at the 16 corners
of the cell, i.e.,
\begin{equation}
w_i = b (\pm l_1 \pm l_2 \pm l_3 \pm l_4 ) ~ ,
\end{equation}
with all 16 possible choices of the signs, and $m_i$ is the number of
minus signs used in each $w_i$.
One can note that similar simple formulas result for this type
of integral in other dimensions as well.\cite{Gilat2d3d}

In the actual application of Eq.\ (\ref{Dcell}), inside the cell, 
the parameter $w$ will range between minimum and maximum values: 
\begin{equation}
w_{\rm max} = b(|l_1|+|l_2|+|l_3|+|l_4|) = - w_{\rm min} ~ .
\end{equation}
These values hold at two opposite corners of the cell.
By Eq.\ (\ref{w-omega}), there will be a corresponding range
of $\omega$ over which this cell contributes to $D(\omega)$.
Then for the general problem where $D$ is desired over a 
large range of $\omega$, what is done is to set a uniform 
grid of $\omega$-values, finding the contribution of each
cell to $D(\omega)$ for each value on the grid, say, $\omega=\omega_n$.
Typically it works well to have the grid fine enough so that
each cell contributes to $D$ at many different $\omega_n$. 
%

\subsection{Numerical calculation of $\Gamma_{\bk}$}
\label{GammaNum}
For the damping problem, Eq.\ (\ref{Gamma}), we have the replacements,
\begin{equation}
f({\bf x})=F(\bq,\bp) ~ , \quad\quad 
g({\bf x})=\omega_{\bk+\bp-\bq}-\omega_{\bp}+\omega_{\bq} ~,
\end{equation}
with the 4D variable ${\bf x}=(\bq,\bp)$.
We want to find $\Gamma(\bk,\omega)$
right at $\omega=\omega_{\bk}$ for some particular wavevector, $\bk$. 
%
%
To accomplish this, we set a range
$\omega_{\rm min} \le \omega \le \omega_{\rm max}$,
and define the grid of values
$\omega_n = \omega_{\rm min} + n (\omega_{\rm max}
-\omega_{\rm min})/N_{\rm grid}$,  using $N_{\rm grid} =500$.
For the integrations over $\bq$ and $\bp$, the cell sizes actually can
be quite large: we used the cell width $2b=2\pi/L$, where $L=41$ 
is a ``system size,'' which gave results consistent with those using 
much smaller cells (larger $L$).  
That is, each axis, $q_x$, $q_y$, $p_x$, $p_y$, was partitioned into 41 cells. 
First, to get good precision, a narrow 
range symmetrical around the spinwave frequency was used:
$\omega_{\rm min}=\omega_{\bk}-0.1 JS$, 
$\omega_{\rm max}=\omega_{\bk}+0.1 JS$.
Second, for comparison, the behavior of the damping function was 
calculated over the whole range of possible spinwave frequencies, i.e.,
$\omega_{\rm min}=0$, $\omega_{\rm max} \approx  \omega_{\bq=(\pi,\pi)}
= 4 \sqrt{2(1+\lambda)} JS $. 
In both of these approaches, interpolation on the $\omega$-grid gives
$\Gamma(\bk,\omega_{\bk})$.
The scheme was found to be very robust, with both approaches 
agreeing to better than one percent precision.
%

We should also note one other important calculational detail.  
Because Eq.\ (\ref{Dcell}) depends on the reciprocal gradient
of $g$,  it is necessary to avoid any points in the $\bq,\bp$-space
where this gradient vanishes.
We accomplished this by taking wavevectors $\bk=(x_k,y_k)\frac{2\pi}{L}$,
(at $L=41$),  and then shifting the $\bq,\bp$-cell
center positions slightly from the usual values:
$\bq=(x_q+\frac{1}{2},y_q+\frac{1}{2})\frac{2\pi}{L}+(\Delta,\Delta)$,
$\bp=(x_p+\frac{1}{2},y_p+\frac{1}{2})\frac{2\pi}{L}+(\Delta,\Delta)$.
Here $x_k, y_k, x_q, y_q$, etc., are integers ranging from 0 to $L-1$,
and we used $\Delta=0.022378$ to shift the $\bq,\bp$-cells 
away from high symmetry points.
The choice of an odd number for $L$ also helps to avoid certain
high symmetry points.
Some slight improvement in precision can be achieved with a larger
value of $L$, however, at the expense of a tremendous increase in 
cpu time, which varies as $4^L$.
At $L=41$, the typical relative errors in $\Gamma(\bk,\omega)$ were found 
to be much less than one percent, when compared to calculations
using $L=81$.

\begin{figure}
\hskip 1.6in
\psfig{figure=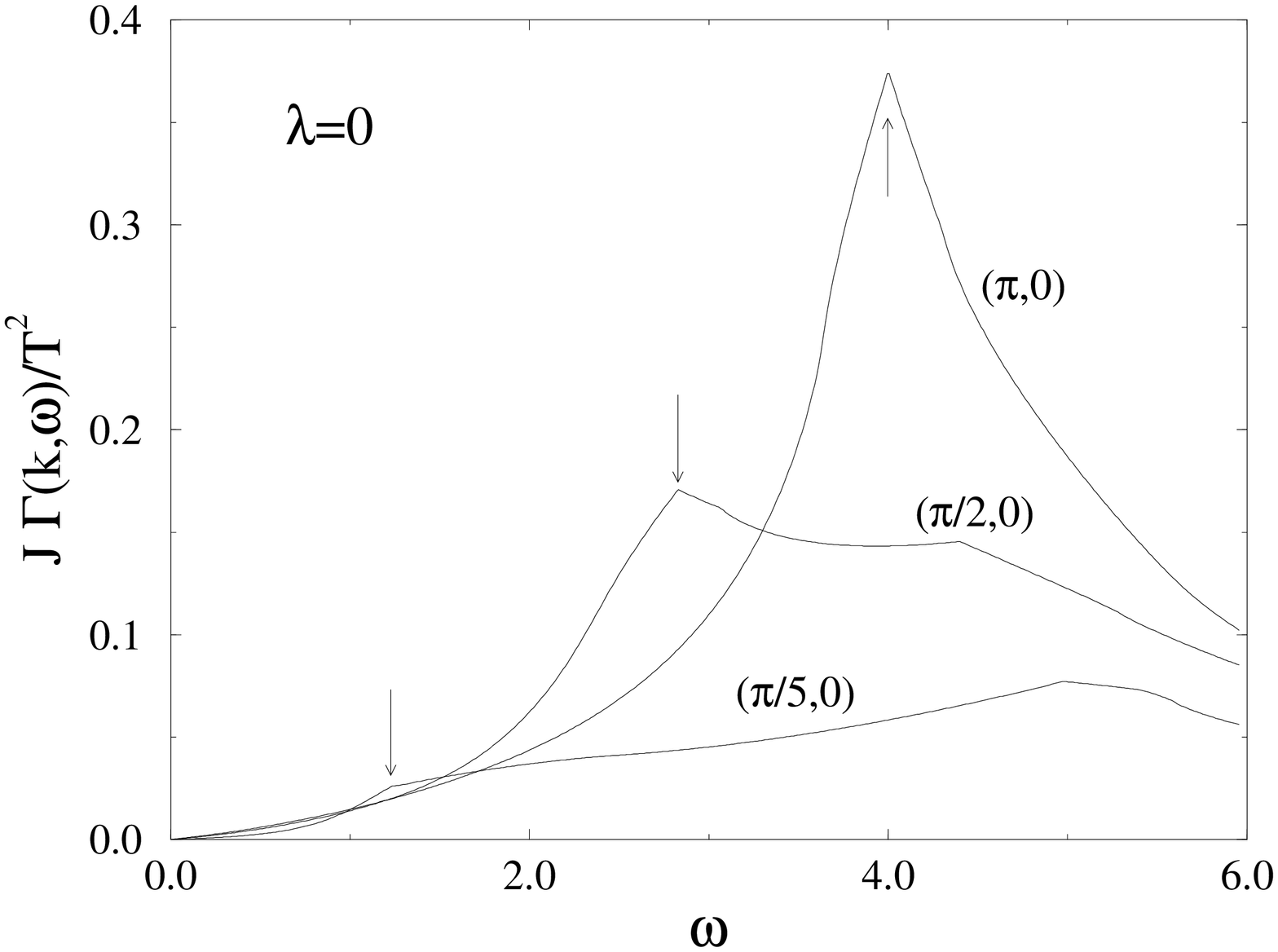,angle=-90.0,width=\pssize pt} 
\psfig{figure=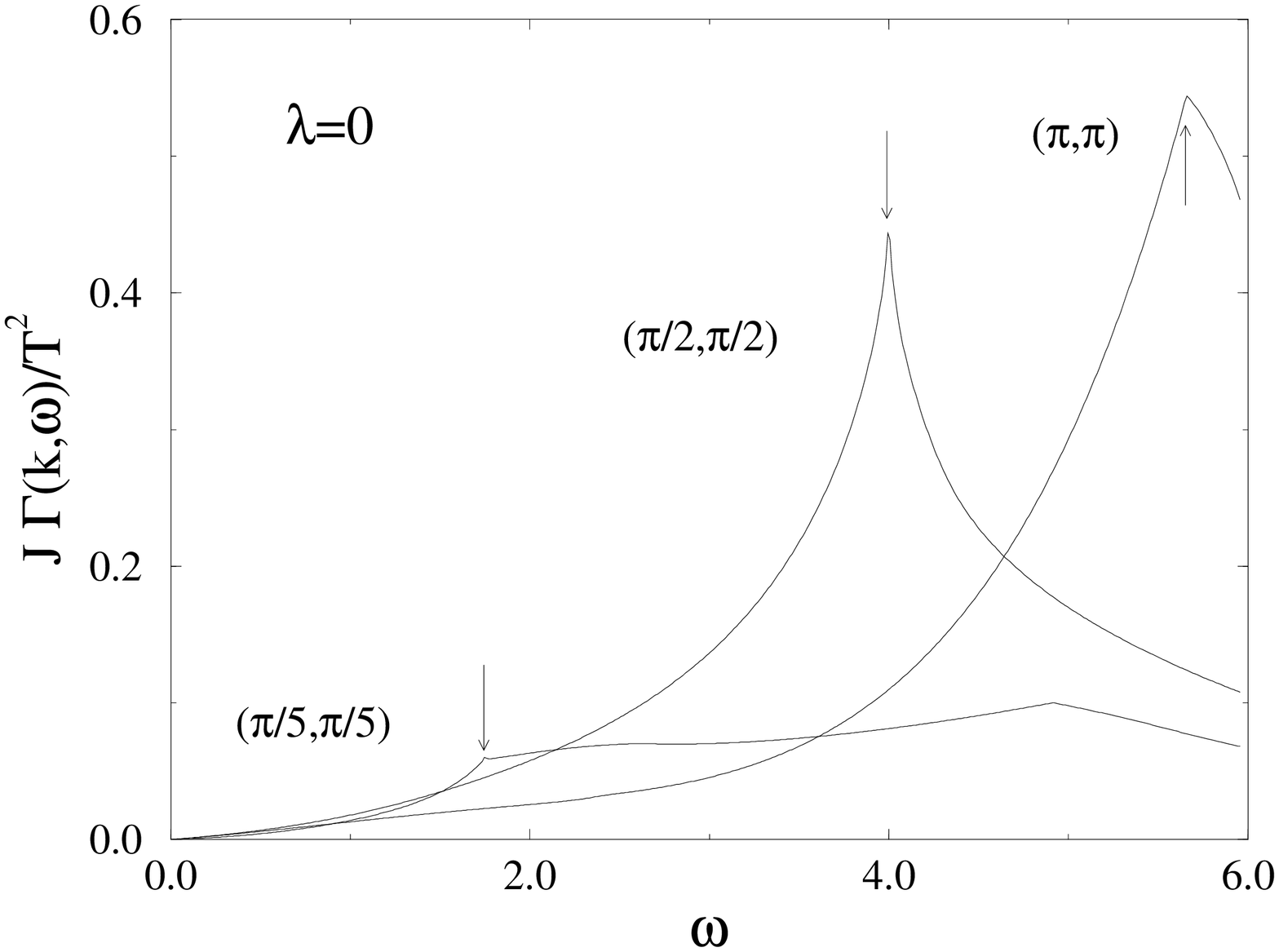,angle=-90.0,width=\pssize pt}
\caption{
\label{Gam_omega-fig}
Typical numerical results for the damping function $\Gamma(\bk,\omega)$
at $\lambda=0$, at various wavevectors as indicated.  
The arrows mark the positions of the spinwave frequency $\omega_{\bk}$,
for each associated wavevector.
}
\end{figure}

\begin{figure}
\hskip 1.6in
\psfig{figure=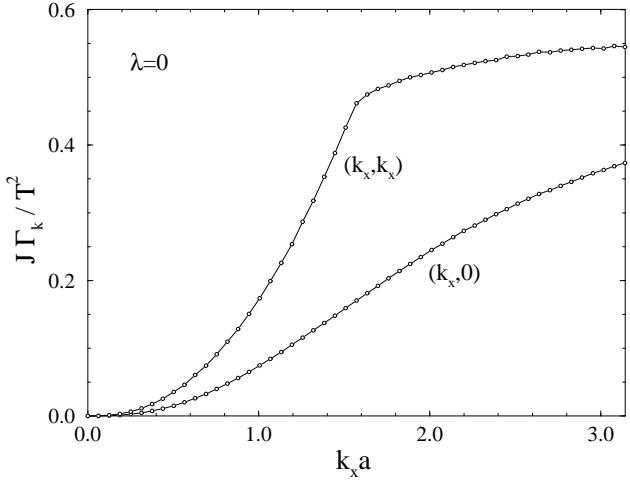,angle=-90.0,width=\pssize pt}
\caption{
\label{Gam_k-fig}
The damping rate $\Gamma_{\bk}$ for $\lambda=0$ 
in the (10) and (11) directions.
}
\end{figure}

Some results for the damping function, which is proportional to
$T^2$,  are shown in Fig.\ \ref{Gam_omega-fig} for  $\lambda=0$.
In general, it is found that $\Gamma(\bk,\omega)$ either has a cusp or
a peak at the spinwave frequency (arrows in Fig.\ \ref{Gam_omega-fig}).
The damping strength increases with $\vert \bk \vert$, as seen in
Fig.\ \ref{Gam_k-fig}.
Along the (11) direction, however, $\Gamma_{\bk}$ tends to saturate 
halfway out in the zone.
Keeping in mind that $\Gamma_{\bk} \propto T^2$, the damping strength
we find here is quite weak, especially when we consider its
effects at $T \ll J$ where the spinwave approximation is valid. 
%

\begin{figure}
\hskip 1.6in
\psfig{figure=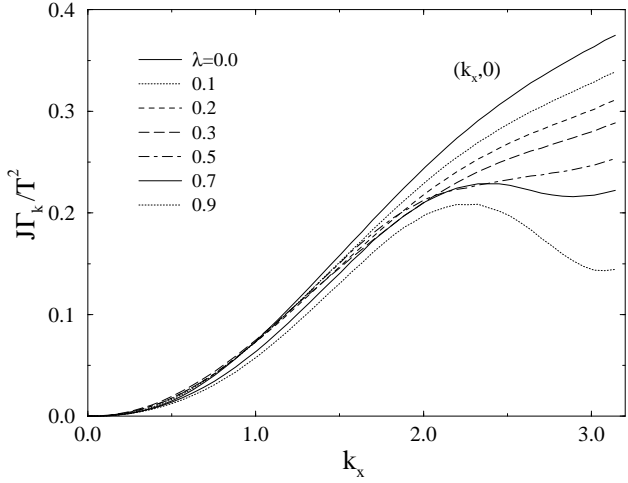,angle=-90.0,width=\pssize pt}
\caption{
\label{Gam_lambda-fig}
The damping rate $\Gamma_{\bk}$ for $0 \le \lambda \le 0.9 $ 
in the (10) direction.
}
\end{figure}

At small nonzero values of the anisotropy parameter $\lambda$, there will
be only a very slight change in $\Gamma_{\bk}$, because $\lambda$ does
not appear explicitly in $H_1$, but only enters the calculation
via the dispersion relation and via the $\alpha$ and $\beta$ functions 
of (\ref{canonical}).
To demonstrate this effect, Fig.\ \ref{Gam_lambda-fig}
shows $\Gamma_{\bk}$ at various values of $\lambda$ 
for $k$ along the (10) direction.
In general, the damping decreases with increasing $\lambda$,
i.e. as the system becomes more isotropic.
%
 
The calculation of the memory function requires the knowledge
of $\Gamma_{\bk}$ over the entire BZ.
We can reduce the region of calculation to $k_x \ge k_y \ge 0$, 
because, by symmetry,  $\Gamma_{\bk}=\Gamma_{-\bk}$,
and $\Gamma_{(k_x,k_y)}=\Gamma_{(k_y,k_x)}$.
In this region $\Gamma_{\bk}$ was calculated on a uniform square grid 
with spacing $\Delta k_x =  \Delta k_y = 2\pi/50$. 
Linear interpolation was used to get values
between these grid points.

\section{The memory function formalism with damping}
\label{Memory}
For the calculation of the dynamic correlation function  $S(q,\omega)$,
we use the memory function formalism due to Reiter and 
Sj\"olander,\cite{RS77} who applied the procedure to the 
isotropic Heisenberg model.
The technique is based on a low-temperature expansion of the memory
function.\cite{Mori65}
The formalism was applied to the 1D isotropic antiferromagnet in
an external field by De Raedt et al\cite{DeRaedt+81} and to a 
1D easy-plane antiferromagnet by Gouv\^ea and Pires.\cite{Gouvea+87} 
More recently, it was also applied to 2D easy-plane ferromagnets
by Menezes et al,\cite{Menezes} although some numerical difficulties 
inhibited numerical evaluation of the analytic formulas.
Some of these problems were surpassed in the work
of Pereira et al.\cite{Pereira,Pereira+00} 
In all of these works, no mode damping was included.
The presence of damping actually relieves the numerical difficulty of
the problem, as we see below.

The dynamic correlation function is defined in terms of the
Fourier transformed spin-fields as
\begin{equation}
\label{Sqw}
S^{\alpha\alpha}(\bq,\omega)=
\frac{1}{2\pi} \int_{-\infty}^{\infty} dt ~ e^{-i\omega t}
\langle S^{\alpha}(\bq,t) S^{\alpha}(-\bq,0) \rangle ~ ,
\end{equation}
where $\alpha$ is any component of $S$, and by symmetry, 
$S^{xx}(\bq,\omega)=S^{yy}(\bq,\omega)$.  
In this and following formulas for correlation functions, we note 
the distinction between continuum (functions of $\bq$)
and discrete (subscripts $\bq$) Fourier transforms, which
are related by factors of $2\pi/L$ for each dimension.\cite{FTNote}
This is necessary so that we can make a careful comparison between the theory
and the MC-SD simulations, {\em including the absolute magnitudes}.
Specifically, the continuum correlation function in (\ref{Sqw}),
which gives a thermodynamic weight over a {\em wavevector range}
$d^{2}q = (2\pi/L)^{2}$, is related to the discrete correlation
at a {\em specific} wavevector $\bq$, using discrete Fourier transforms 
as defined in (\ref{Fourier}),  by\cite{FourierNote}
\begin{equation}
\langle S^{\alpha}(\bq,t) S^{\alpha}(-\bq,0) \rangle =
\frac{1}{(2\pi)^{2}} 
\langle S_{\bq}^{\alpha}(t) S_{-\bq}^{\alpha}(0) \rangle ~ .
\end{equation}
In the analytic calculations, many intermediate quantities are
discrete functions, however, for the comparison with the MC-SD data,
both results will be converted and displayed as continuum quantities. 

Due to the symmetry in the $xy$-plane, the calculation is carried
out by making averages of the rotationally symmetric quantity, 
$S_{\bq}^{\perp}=S_{\bq}^x + S_{\bq}^y$,  
with $S^{xx}(\bq,\omega)=\frac{1}{2}S^{\perp\perp}(\bq,\omega)$.
Then in the memory function calculation, the continuum
dynamical structure function
can be written in terms of a second-order memory function, 
$\Sigma_{\bq}^{\perp}(\omega)$, to be described below, according to
\begin{eqnarray}
\label{Sqw-memory}
&S&^{\perp\perp}(\bq,\omega) = \nonumber \\
&&  \frac{ (2\pi)^{-2}  \pi^{-1}
\langle LS_{\bq}^{\perp},LS_{-\bq}^{\perp} \rangle 
~{\rm Im} ~ \Sigma_{\bq}^{\perp}(\omega)}
{\left[\omega^2 - \langle \omega^2 \rangle_{\bq}^{\perp} 
+\omega ~{\rm Re} ~ \Sigma_{\bq}^{\perp}(\omega) \right]^2
+ \left[ \omega ~{\rm Im} ~ \Sigma_{\bq}^{\perp}(\omega) \right]^2 } ~ ,
\end{eqnarray}
where $L=-i ~ d/dt$ is the Liouville operator,  
and the second moment $\langle \omega^2 \rangle_{\bq}^{\perp}$
is determined from a ratio,
\begin{equation}
\label{moment2}
\langle \omega^2 \rangle_{\bq}^{\perp} =
\frac{\langle LS_{\bq}^{\perp},LS_{-\bq}^{\perp}\rangle}
{\langle S_{\bq}^{\perp},S_{-\bq}^{\perp}\rangle} ~ .
\end{equation}
The location of the spinwave peak that results from 
Eq.\ (\ref{Sqw-memory}) is very sensitive to the value of 
$\langle \omega^2 \rangle_{\bq}^{\perp}$,
therefore we are especially concerned to get an 
accurate evaluation of it.
The denominator of (\ref{moment2}) is a static correlation function, 
whose value will be taken from the Monte Carlo simulations, although for 
low temperatures it can be evaluated approximately as 
\begin{mathletters}
\label{Sqperp}
\begin{equation}
\langle S_{\bq}^{\perp} S_{-\bq}^{\perp} \rangle \approx 
\left(1-\langle z_{\bn}^2 \rangle \right) 
\frac{T}{4J} 
(1-\gamma_{\bq})^{-1} ~ ,
\end{equation}
\begin{equation}
\langle z_{\bn}^2 \rangle \approx  \frac{T}{4J}  ~ ,
\end{equation}
\end{mathletters}
The numerator of (\ref{moment2}) has been 
shown\cite{Menezes,Pereira} to have a leading
low-temperature dependence, also related to static nearest 
neighbor correlations,
\begin{eqnarray}
\label{LSLS}
\langle &L& S_{\bq}^{\perp},LS_{-\bq}^{\perp} \rangle = \nonumber \\
&& 4JT(1-\lambda\gamma_{\bq}) 
\langle S_{a\hat{x}}^{\perp} S_{0}^{\perp} \rangle
-8JT(\gamma_{\bq}-\lambda) \langle S_{a\hat{x}}^{z} S_{0}^{z} \rangle ~ .
\end{eqnarray}
The correlation $\langle S_{a\hat{x}}^{z} S_{0}^{z} \rangle \propto
\lambda T$, therefore its contribution can be ignored for small $\lambda$. 
There being two bonds per lattice site, the in-plane near-neighbor 
correlation is related to the average energy per bond by
\begin{equation}
\label{S0Sa-EN}
\left\langle \frac{E}{2JN} \right\rangle = 
S^2 - \langle \vec{S}_{a\hat{x}} \cdot \vec{S}_{0} \rangle \approx 
S^2 - \langle S_{a\hat{x}}^{\perp} S_{0}^{\perp} \rangle ~ .
\end{equation}
This correlation can also be obtained from the MC simulations,
although a good low-T approximation is
\begin{equation}
\label{S0Sa-theory}
\langle S_{a\hat{x}}^{\perp} S_{0}^{\perp} \rangle \approx
\left(1-\frac{T}{4J} \right) e^{-1.273\frac{T}{4J}} ~ .
\end{equation}
A comparison of the analytic and MC static parameters will be
given below.

To the leading order in temperature, the time-dependent memory 
function has been shown to be given from
\begin{mathletters}
\begin{equation}
\Sigma_{\bq}^{\alpha}(t) = 
- \langle LS_{\bq}^{\alpha} LS_{-\bq}^{\alpha} \rangle^{-1}
M_{\bq}^{\alpha}(t) ~ ,
\end{equation}
\begin{equation}
M_{\bq}^{\alpha}(t) = 
\langle QL^2 S_{\bq}^{\alpha} e^{-iLt} QL^2 S_{-\bq}^{\alpha} \rangle ~ , 
\end{equation}
\end{mathletters}
where $Q$ is projection onto the non-secular variables, see Ref.\
\onlinecite{RS77}, whose action is equivalent to
\begin{equation}
QL^2 S_{\bq}^{\alpha} = (L^2-\langle \omega^2 \rangle_{\bq}^{\alpha})
S_{\bq}^{\alpha} ~ .
\end{equation}
%
%
Using expressions (\ref{mapping}) expanded to second order in
$\phi_{\bn}$, $z_{\bn}$, and (\ref{canonical}),
it is straightforward to write the spins $S_{\bn}$ in terms of 
the creation/annihilation operators. 
In the leading order in temperature, the resulting expressions  
involve expectation values of products of four creation/annihilation
operators.  
The damping is included in the time evolution according to,
\begin{equation}
a_{\bq}^{\dagger}(t)=a_{\bq}^{\dagger}(0)e^{(i\omega-\Gamma_{\bq})t}
\quad , \quad
a_{\bq}(t)=a_{\bq}(0)e^{(-i\omega-\Gamma_{\bq})t} ~ .
\end{equation}
Then, there results, after some manipulations, expressions
to give the memory function,
\begin{eqnarray}
\label{M-memory}
 M_{\bq}^{\perp}(t) = \frac{1}{N} \sum_{\bp} & \Bigl\{ &
 ~ W_{+}(\bq,\bp) e^{-\Gamma_{+}(\bq,\bp)t} \cos\Omega_{+}(\bq,\bp)t 
\nonumber \\
&& + 
W_{-}(\bq,\bp) e^{-\Gamma_{-}(\bq,\bp)t} \cos\Omega_{-}(\bq,\bp)t 
\Bigr\} ~ .
\end{eqnarray}
The two terms correspond to spinwave sum ($\Omega_{+}$) and 
difference ($\Omega_{-}$)  processes, with weights
\begin{equation}
\label{weights}
W_{\pm}(\bq,\bp) = n(\omega_{\frac{\bq}{2}+\bp}) n(\omega_{\frac{\bq}{2}-\bp}) 
(\tilde{s} \mp \tilde{t} ~ )^2 ~ ,
\end{equation}
where
\begin{mathletters}
\begin{eqnarray}
\tilde{s} = S ~ & \beta_{\frac{\bq}{2}+\bp} \beta_{\frac{\bq}{2}-\bp} 
\Bigl\{ & \omega_{\bq}^{2} - (4JS)^{2} \times \nonumber \\
&& \Bigl[ ~ 
(1-\gamma_{\frac{\bq}{2}+\bp})
(\gamma_{\frac{\bq}{2}-\bp}-\lambda\gamma_{\frac{\bq}{2}+\bp})
\nonumber \\
&& + 
(1-\gamma_{\frac{\bq}{2}-\bp})
(\gamma_{\frac{\bq}{2}+\bp}-\lambda\gamma_{\frac{\bq}{2}-\bp}) 
\Bigr]
\Bigr\} ~ ,
\end{eqnarray}
\begin{eqnarray}
\tilde{t} = S ~ & \alpha_{\frac{q}{2}+p} \alpha_{\frac{q}{2}-p} 
\Bigl\{ & \omega_{q}^{2} + (4JS)^{2} \times \nonumber \\
&& \Bigl[ ~ 
(1-\lambda\gamma_{\frac{q}{2}-p})
(\gamma_{\frac{q}{2}-p}-\lambda\gamma_{\frac{q}{2}+p}) 
\nonumber \\
&& + 
(1-\lambda\gamma_{\frac{q}{2}+p})
(\gamma_{\frac{q}{2}+p}-\lambda\gamma_{\frac{q}{2}-p})
\Bigr] \Bigr\} ~ .  
\end{eqnarray} 
\end{mathletters} 
The frequencies and damping rates of these processes are
\begin{mathletters}
\begin{equation}
\Omega_{\pm}(\bq,\bp) = 
\omega_{\frac{\bq}{2}+\bp} \pm \omega_{\frac{\bq}{2}-\bp} ~ ,
\end{equation}
\begin{equation}
\Gamma_{\pm}(\bq,\bp) = 
\Gamma_{\frac{\bq}{2}+\bp} + \Gamma_{\frac{\bq}{2}-\bp} ~ .
\end{equation}
\end{mathletters}
%

Finally, the frequency-dependent memory function is given by the
one-sided Fourier-Laplace transform,
\begin{equation}
\Sigma_{\bq}^{\perp}(\omega) = 
-i \int_{0}^{\infty} dt ~ e^{i\omega t} ~ \Sigma_{\bq}^{\perp}(t) ~ .
\end{equation}
From this expression, Eq.\ (\ref{M-memory}), and equations following, 
we obtain a sum over all the sum/difference processes,
\begin{eqnarray}
\label{Sigma_sum}
&\Sigma&_{\bq}^{\perp}(\omega) =  
 \frac{-1\ }{2N} \langle LS_{\bq}^{\perp} LS_{-\bq}^{\perp} \rangle^{-1}
\sum_{\bp} 
\nonumber \\
&& 
\Bigl\{  
W_{+} \Bigl[ 
 \frac{\omega+\Omega_{+}-i\Gamma_{+}}{(\omega+\Omega_{+})^{2}+\Gamma_{+}^{2}}
+\frac{\omega-\Omega_{+}-i\Gamma_{+}}{(\omega-\Omega_{+})^{2}+\Gamma_{+}^{2}}
\Bigr] 
\nonumber \\
&& 
+W_{-} \Bigl[ 
 \frac{\omega+\Omega_{-}-i\Gamma_{-}}{(\omega+\Omega_{-})^{2}+\Gamma_{-}^{2}}
+\frac{\omega-\Omega_{-}-i\Gamma_{-}}{(\omega-\Omega_{-})^{2}+\Gamma_{-}^{2}}
\Bigr]
\Bigr\} ~ .
\end{eqnarray}
where the $(\bq,\bp)$ dependencies of $W_{\pm}, \Omega_{\pm}$, 
and $\Gamma_{\pm}$ have been suppressed.
This is the expression used to evaluate $\Sigma_{\bq}^{\perp}$,
and thus the dynamic structure function, via Eq.\ (\ref{Sqw-memory}).
%

To evaluate (\ref{Sigma_sum}) on a discrete $L\times L$ lattice, 
the wavevector $\bq$ must be one of the allowed values,
\begin{equation}
\bq = (x_q,y_q) (2\pi/L) ~ ,
\end{equation}
where $x_q$ and $y_q$ are integers.
It is also necessary to make the summations over $\bp$ such that
the vectors $\frac{\bq}{2}\pm \bp$ are both given by integers times $2\pi/L$.
Thus, it is convenient to change from the summation over $\bp$ 
to a summation over a new variable $\bk$ defined by
\begin{equation}
\label{k-discrete}
\bk \equiv \frac{\bq}{2}+\bp = (x_k,y_k) (2\pi/L),
\end{equation}
where $x_k$ and $y_k$ are integers.
Then, we also have $\frac{\bq}{2}-\bp = \bq-\bk$ equal to an allowed 
wavevector.
This insures that the sums will be over spinwaves that are
allowed on the given finite lattice.
This approach is essential for calculation of the finite size 
dynamical effects.
On the other hand, for the infinite sized system, no variable change is
needed, and there is the usual replacement,
$\frac{1}{N}\sum_{\bp}\rightarrow (2\pi)^{-2} \int d^{2}p$.

Note that in the limit of zero damping, Eq.\ (\ref{Sigma_sum})
is inapplicable, except for the continuum limit (i.e., $L\rightarrow\infty$).
The resulting continuum integrals for the infinite system are singular, 
and the analytic relation (\ref{delta-relation}) must be applied,
leading to 
\begin{eqnarray}
\label{Sigma_int}
&\Sigma&_{\bq}^{\perp}(\omega) =  
\frac{1}{2(2\pi)^{2}} 
\langle LS_{\bq}^{\perp} LS_{-\bq}^{\perp} \rangle^{-1} \times
\nonumber \\
&& 
\Bigl\{  
~ P \int d^{2}p ~ W_{+} 
\Bigl[ \frac{1}{\omega+\Omega_{+}} +\frac{1}{\omega-\Omega_{+}} \Bigr] 
\nonumber \\
&& 
+ P \int d^{2}p ~ W_{-} 
\Bigl[ \frac{1}{\omega+\Omega_{-}} +\frac{1}{\omega-\Omega_{-}} \Bigr] 
\nonumber \\
&& 
-i \pi  \int d^{2}p ~ W_{+} 
\Bigl[ \delta(\omega+\Omega_{+}) +\delta(\omega-\Omega_{+}) \Bigr]
\nonumber \\
&& 
-i \pi  \int d^{2}p ~ W_{-} 
\Bigl[ \delta(\omega+\Omega_{-}) +\delta(\omega-\Omega_{-}) \Bigr]
\Bigr\} ~ .
\end{eqnarray}
Indeed, in the very low temperature range ($T < 0.2 J$) where
the damping is extremely weak, it may be more efficient to use 
this expression rather than Eq.\ (\ref{Sigma_sum}),
which has poorer convergence with $L$.
Eq.\ (\ref{Sigma_int}) was applied by Pereira et al.\cite{Pereira,Pereira+00}; 
the imaginary part can be evaluated by techniques described 
here (Sec.\ \protect\ref{Integration}),
and the principal-valued real part can be calculated by an 
extension\cite{Wysin0} of techniques used in 1D.\cite{Gouvea0}
In Ref.\ \onlinecite{Pereira+00} the last expression was found to 
give finite amplitude spinwave peaks, without damping present.
%

Before presenting the results obtained via the memory function
approach, the key aspects of the Monte Carlo and spin-dynamics
simulations are summarized.

\section{Monte Carlo and spin-dynamics}
\label{MC+SD}
To test the validity of the memory function technique, we 
compare it with numerical simulations on $L \times L$ square 
lattices ($L=128$, $\lambda=0$)
using Monte-Carlo and spin-dynamics (SD) simulations, which
include effects due to all thermodynamically allowed excitations.
As mentioned above, we also use some data for static correlations
from the Monte Carlo calculations as input to the memory function
dynamical calculations.
The techniques used here have been described in Ref.\ 
\onlinecite{GouveaWysin} and are based on the simulation methods
developed in Ref.\ \onlinecite{MC-SD-Refs}.
%

In the MC calculation, we applied a combination of 
Metropolis single-spin moves and over-relaxation moves 
that modify all three spin components,
and in addition, Wolff single-cluster operations\cite{Wolff}
that modify {\em only} the $xy$ spin components.
The over-relaxation and cluster moves are important at
low temperatures, where the $xy$ spin components tend to freeze
and single spin moves become inefficient.
%

In the combined MC-SD simulations, a set of MC generated states
from thermodynamic equilibrium at temperature $T$ are used as
initial conditions for numerical integrations in time via
a fourth order Runge-Kutta scheme.
Specifically, the first 4000 MC steps were used for equilibrating the
system.
Then, every 400 MC steps a state was taken to initiate the SD integration.
This was repeated 500 times; the displayed $S(\bq,\omega)$ data are
averages over the 500 time integrations.
The total MC sequence was then $2\times 10^5$ steps; every 25 steps
a state was used to calculate averages of the static correlation functions.
%

In the time integration, the basic Runge-Kutta time step was 
$\Delta t = 0.03 (JS)^{-1}$, and every 11 time steps, data samples
(for $S_{\bq}^{x}(t)$, etc.)  were saved for $\bq$'s in the (10), (01), 
and (11) directions only (due to machine memory restrictions).
A total of $2^{12}$ data samples were saved, followed by applying
a fast-Fourier-transform to obtain $S^{xx}(\bq,\omega)$.
The total time of integration of $t_{\rm max}=1351.68 (JS)^{-1}$ is 
large enough so that no smoothing window is necessary for the data 
generation.
Then the frequency resolution of these results is good, 
$\Delta\omega = 2\pi/t_{\rm max} = 0.0046484 JS$, so that any
fine details in the low temperature results should be visible. 
For the lowest temperatures ($T \le 0.3 J$), where 
we wanted to see the complex fine details due to the finite 
size effects, some simulations were
made saving data samples every 28 times steps, leading to a total
integration time $t_{\rm max}=3440.64 (JS)^{-1}$, and frequency
resolution $\Delta\omega = 2\pi/t_{\rm max} = 0.0018262 JS$.

\section{Numerical results}
\label{Results}
First, a comparison of the low-$T$ formulas and static MC results 
needed as input to the memory function formalism is presented.
Then dynamical results from the memory function and from MD-SD
will be presented for large systems, and finally, for smaller systems,
where the finite size features will be analyzed.
 
\subsection{Statics}
\begin{figure}
\hskip 1.6in
\psfig{figure=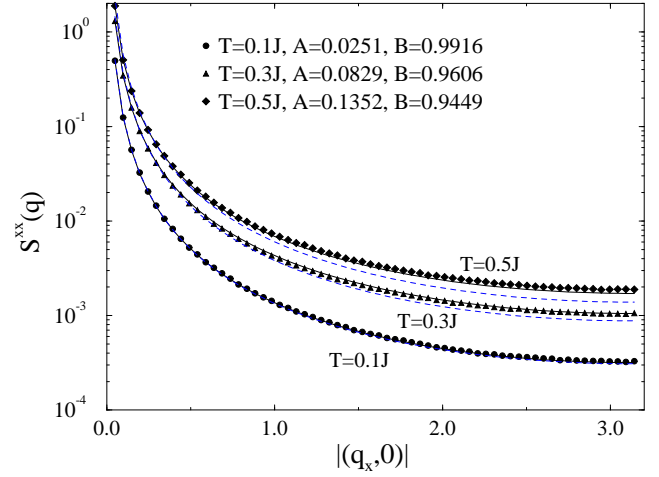,angle=-90.0,width=\pssize pt}
\caption{
\label{Sq-fig}
Static correlation function for $\lambda=0$, calculated from MC
simulations for $L=128$ (symbols), with fits to formula 
(\protect\ref{Sq-fit}) (solid curves), compared to the 
low-temperature expression (\protect\ref{Sqperp}) (dashed curves).
}
\end{figure}

The continuum static structure function,
\begin{equation}
S^{xx}(\bq)=\frac{1}{2(2\pi)^2} 
\langle S_{\bq}^{\perp} S_{-\bq}^{\perp} \rangle ~ ,
\end{equation}
was measured directly in the MC simulations.
A comparison with the results based on the low-temperature expression
(\ref{Sqperp}) is given in Fig.\ \ref{Sq-fig}. 
There we also show a fit of the MC data to a form similar to the 
analytic expression, but with a fitting coefficient ($A$) and power ($B$),
\begin{equation}
\label{Sq-fit}
\langle S_{\bq}^{\perp} S_{-\bq}^{\perp} \rangle_{\rm fitted} =
A (1-\gamma_{\bq})^{-B} ~ .
\end{equation}
From the fits, the parameter $A$ is somewhat larger than the
low-temperature limit, $A=T/(4J)$.
This can be expected due to the renormalization of the effective
$J$ to lower values with temperature.
The power $B$ falls with increasing $T$, which must also be due
to anharmonic effects.
There is little difference between the low-T expression and the
MC data at the lowest temperature.
The deviations between the MC and low-T theory increase substantially
by $T=0.5 J$, especially at higher wavevectors.
Note that the fits are only shown for demonstration; we used the 
MC data itself as inputs to the memory function calculations. 

\begin{figure}
\hskip 1.6in
\psfig{figure=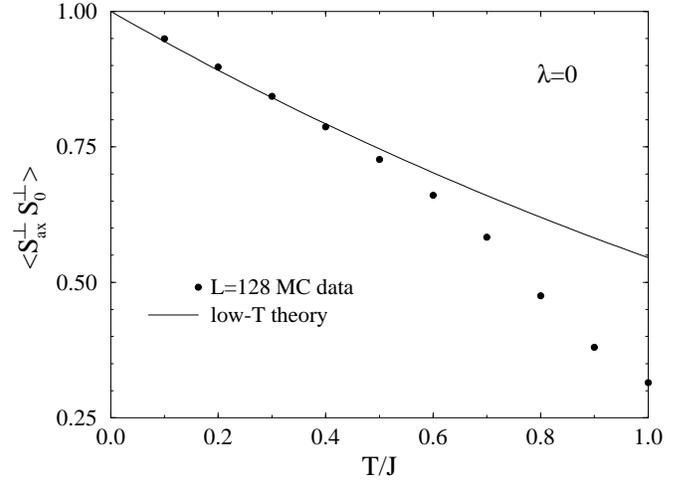,angle=-90.0,width=\pssize pt}
\caption{
\label{S0Sa-fig}
Static nearest neighbor correlation function for $\lambda=0$, 
calculated from MC simulations for $L=128$ using Eq.\ 
(\protect\ref{S0Sa-EN}) (symbols), compared to the 
low-temperature expression 
(\protect\ref{S0Sa-theory}) (solid curve).
}
\end{figure}

The other static input to the memory function calculation, for
determining the second moment, is the quantity 
$\langle LS_{\bq}^{\perp} LS_{-\bq}^{\perp} \rangle$, which is related
to the nearest neighbor correlation 
$\langle S_{a\hat{x}}^{\perp} S_{0}^{\perp} \rangle$ by Eq.\ 
(\ref{LSLS}).
Estimates of that latter from the MC evaluation of the average
energy per spin [Eq.\ (\ref{S0Sa-EN})] are compared to the theoretical
expression (\ref{S0Sa-theory}) in Fig.\ \ref{S0Sa-fig}.
The analytical expression is seen to give results consistent with 
the MC data up to $T\approx 0.5 J$.
%
%

\begin{figure}
\hskip 1.6in
\psfig{figure=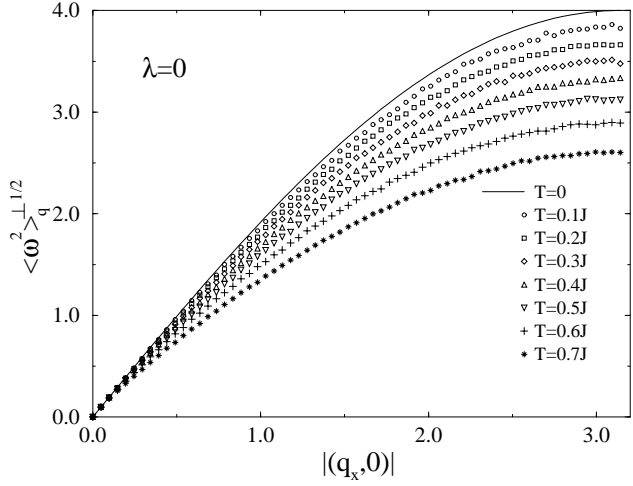,angle=-90.0,width=\pssize pt}
\caption{
\label{moment2-fig}
The second moment vs. $\bq$ for $\lambda=0$, at the temperatures
indicated, using the data from $L=128$ MC simulations, compared
with the zero-temperature dispersion relation (solid curve).
}
\end{figure}

When combined to give the second moment, the MC data produces 
results as shown in Fig.\ \ref{moment2-fig}.
It is essentially this quantity that has the strongest influence
on the locations of the spinwave peaks in $S(\bq,\omega)$.
These are the data used as input to the memory function calculation.

\subsection{Dynamics: Memory Function compared with MC-SD}
We evaluate $\Sigma_{\bq}^{\perp}(\omega)$ by using the sum over
discrete wavevectors $\bp$ as displayed in Eq.\ (\ref{Sigma_sum}),
and then apply Eq.\ (\ref{Sqw-memory}) to determine 
$S^{xx}(\bq,\omega)=\frac{1}{2}S^{\perp\perp}(\bq,\omega)$.
If $L$ is taken large enough, then the sum should be a good
approximation to the infinite size limit. 
In general, we can see whether larger $L$ is necessary 
by the presence of finite size features in the data for
$\Sigma_{\bq}^{\perp}$ and $S^{xx}(\bq,\omega)$.
Typically we used $L=1024$ to give a reasonable approximation
to the continuum limit for $T \ge 0.3 J$ and larger $\bq$-values; 
at lower temperatures or wavevectors even larger $L$ is necessary 
to smooth out all the finite size features, due to the very weak damping.
%

\begin{figure}
\psfig{figure=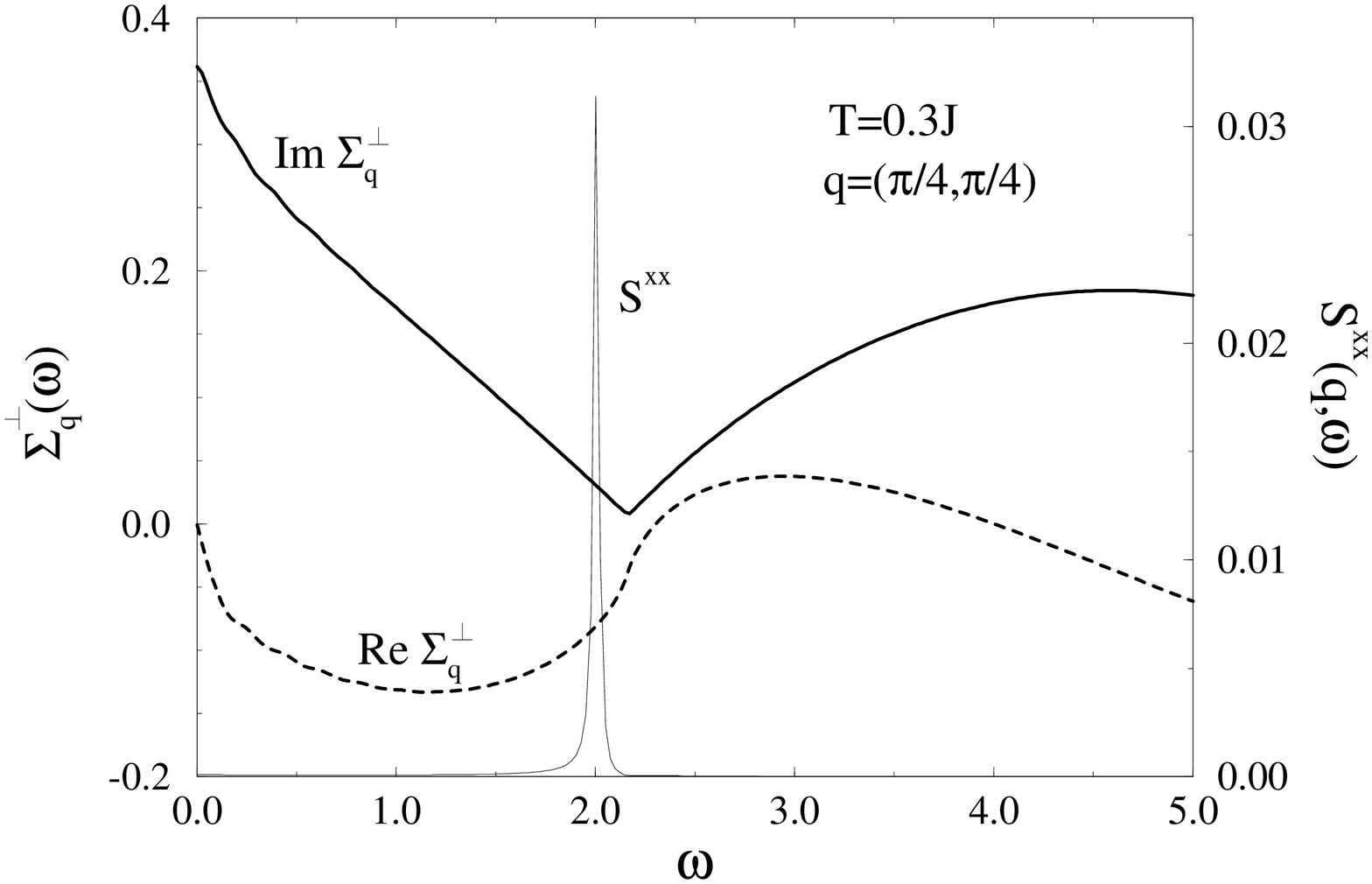,angle=-90.0,width=\pssize pt}
\caption{
\label{SigT3qpi4pi4-fig}
Real and imaginary parts of the memory function 
$\Sigma_{\bq}^{\perp}(\omega)$,  calculated using $L=2048$ for
$T=0.3 J$, $\bq=(\pi/4,\pi/4)$, and the resulting 
dynamical structure function $S^{xx}(\bq,\omega)$. 
}
\end{figure}

\begin{figure}
\hskip 1.6in
\psfig{figure=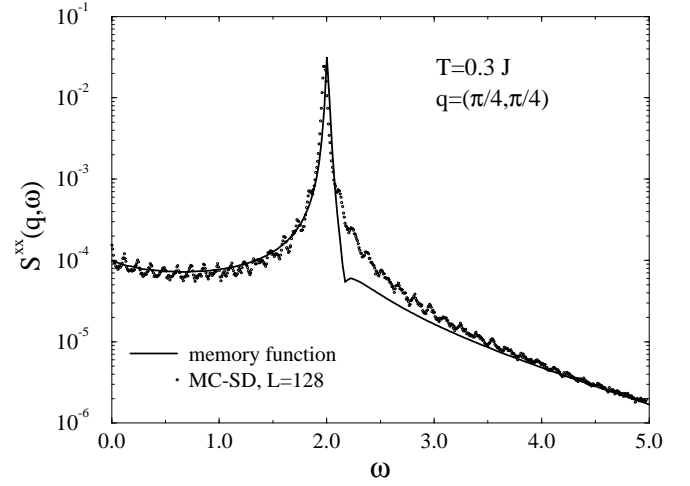,angle=-90.0,width=\pssize pt}
\caption{
\label{SxxT3qpi4pi4-fig}
The dynamical structure function obtained from the memory
function of Fig.\ \protect\ref{SigT3qpi4pi4-fig} (solid curve), 
calculated using $L=2048$ for $T=0.3 J$, $\bq=(\pi/4,\pi/4)$. 
compared with MC-SD data (symbols) for the $L=128$ system.
The frequency resolution of the SD data here and in
Figs.\ \protect\ref{SxxT1-fig}, \protect\ref{SxxT3-fig}, 
\protect\ref{SxxT5-fig} is $\Delta\omega = 0.0046484 JS$.
}
\end{figure}

In Fig.\ \ref{SigT3qpi4pi4-fig}, some typical results for the real 
and imaginary parts of $\Sigma_{\bq}^{\perp}(\omega)$ are shown for 
$\lambda=0$, $T=0.3J$, and $\bq=(\pi/4,\pi/4)$, where $L=2048$ was
used in the sum.
The resulting dynamical structure function is compared to the
corresponding MC-SD data for $L=128$ in Fig.\ \ref{SxxT3qpi4pi4-fig}.  
At the (zero-$T$) spinwave frequency $\omega_{\bq}=2.16478 JS$, 
Im $\Sigma_{\bq}^{\perp}(\omega)$ achieves its minimum value,
which is not zero due to the presence of the damping, while 
Re $\Sigma_{\bq}^{\perp}(\omega)$ has a small nonzero value 
there as well.  
The spinwave peak in the memory data occurs at 
$\omega=\omega_{\rm peak}=2.005 JS$, which is just slightly 
greater than the associated value of the second moment, 
$\omega_{\bq}^{\perp} \equiv 
\sqrt{\langle \omega^2 \rangle_{\bq}^{\perp}} = 1.9648 JS$. 
The spinwave peak position will occur where the denominator
of Eq.\ (\ref{Sqw-memory}) achieves its minimum value, which gives,
to a good approximation,
\begin{equation}
\omega_{\rm peak} =  \omega_{\bq}^{\perp}  
\sqrt{1- {\rm Re}
\Sigma_{\bq}^{\perp}(\omega_{\bq}^{\perp})/\omega_{\bq}^{\perp} } ~ .
\end{equation}
For this case, the memory function data give 
${\rm Re}\Sigma_{\bq}^{\perp}(\omega_{\bq}^{\perp}) \approx -0.081 JS$,  
which is consistent with this relation.
The peak position is very sensitive to the value of the second moment,
which is why we have tried to estimate it as accurately as 
possible from the static MC data.
Still, the MC-SD data show the spinwave peak closer to $\omega=1.98 JS$, 
slightly lower than the memory function prediction.
Using Eq.\ (\ref{Sqw-memory}), the spinwave peak height is proportional
to $[{\rm Im}\Sigma(\omega_{\rm peak})]^{-1}$, while the peak width
is approximately equal to ${\rm Im}\Sigma(\omega_{\rm peak})$.
When compared to the MC-SD data, the memory function calculation 
usually overestimates the peak height and underestimates the peak width,
although both are in reasonable agreement with the MC-SD data.
In fact, these errors tend to be larger at lower temperature.

\begin{figure}
\hskip 1.6in
\psfig{figure=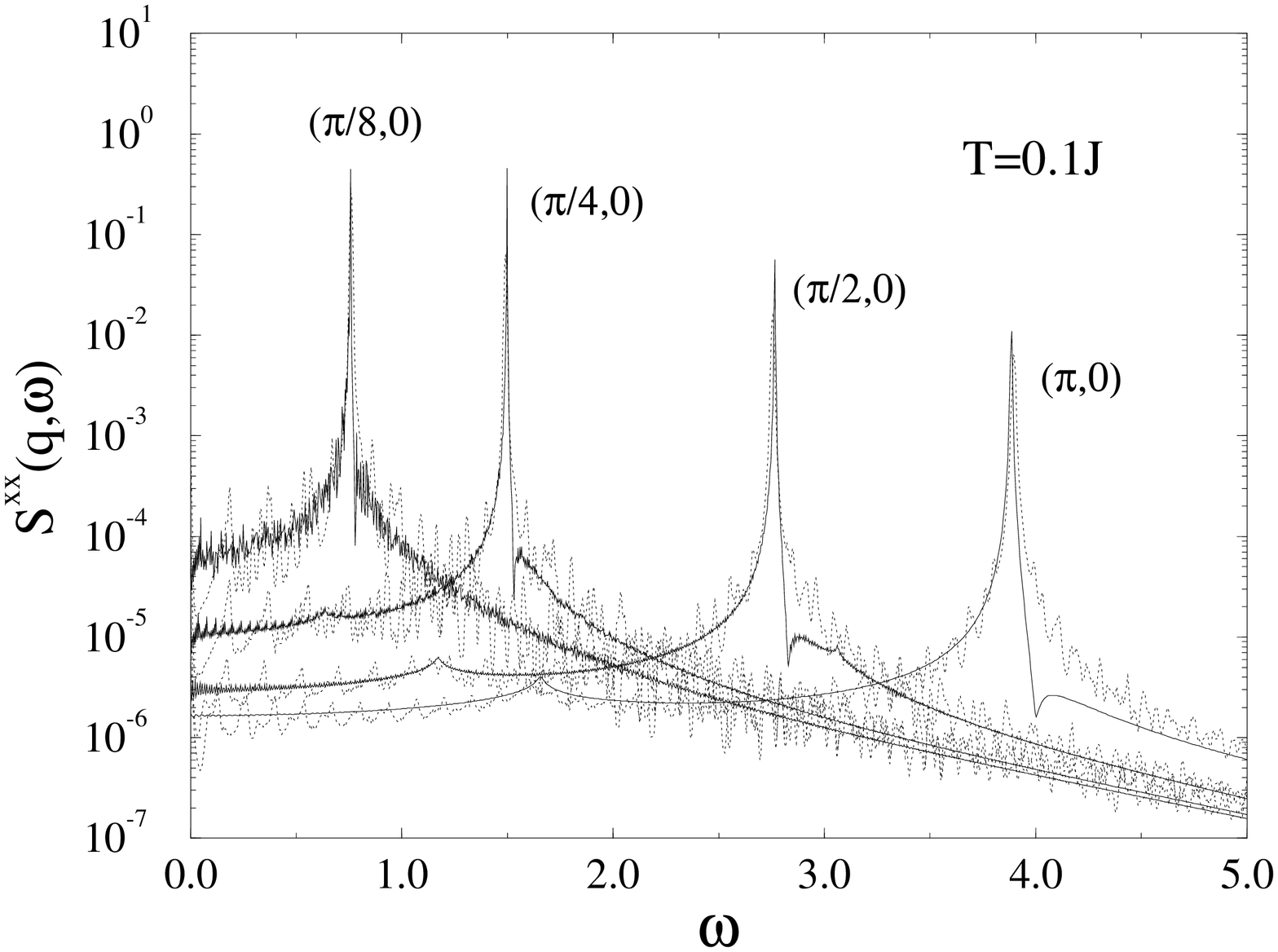,angle=-90.0,width=\pssize pt}
\hskip 1.6in
\psfig{figure=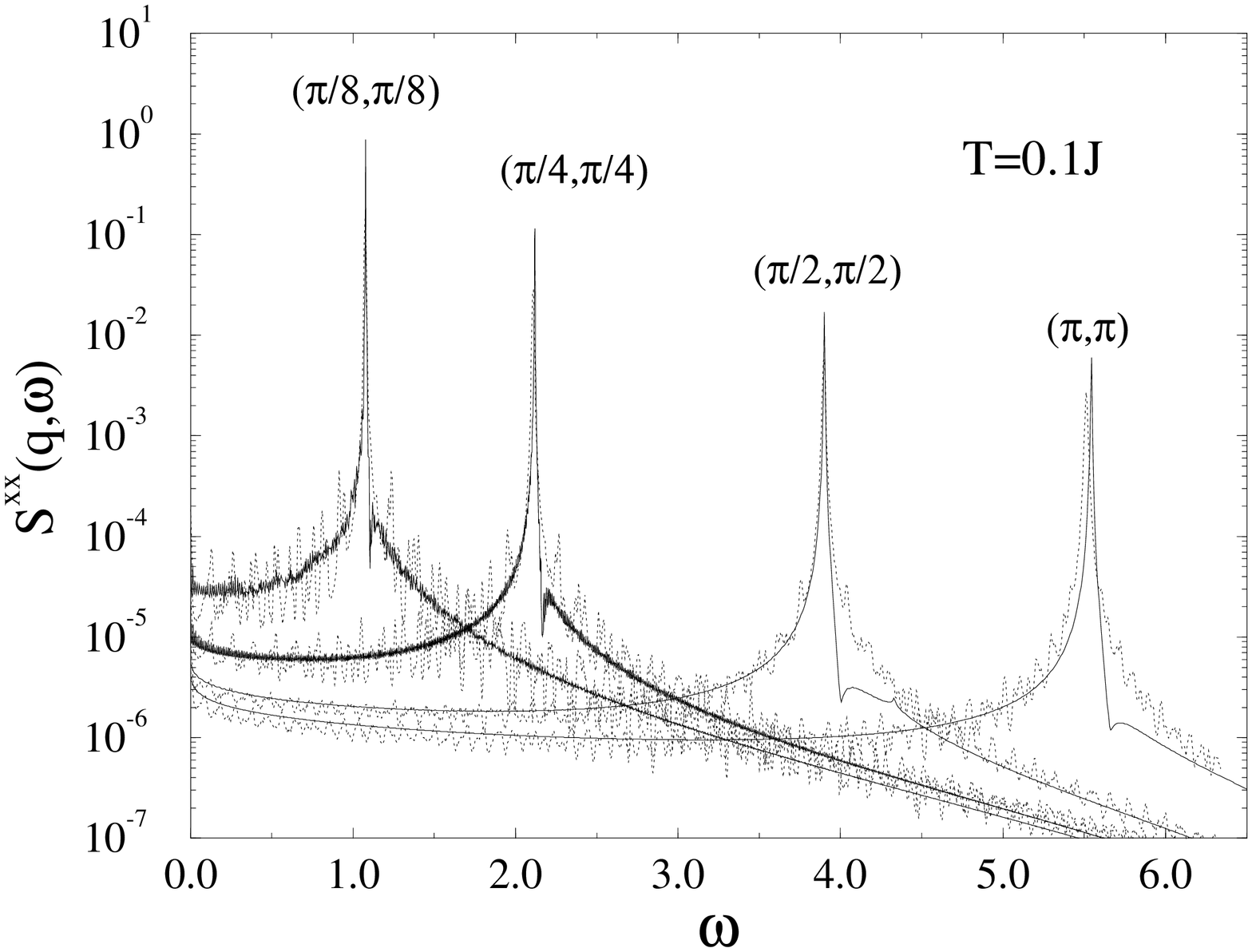,angle=-90.0,width=\pssize pt}
\caption{
\label{SxxT1-fig}
The $T=0.1 J$ dynamical structure function obtained from the memory
function calculations (solid, using $L = 2048$--$4096$) 
and MC-SD simulations (dotted) for the $L=128$ system,
at wavevectors indicated in parenthesis.  The memory data 
and the MC-SD data are both smeared by finite-size effects, 
even for these large $L$.  
}
\end{figure}

\begin{figure}
\hskip 1.6in
\psfig{figure=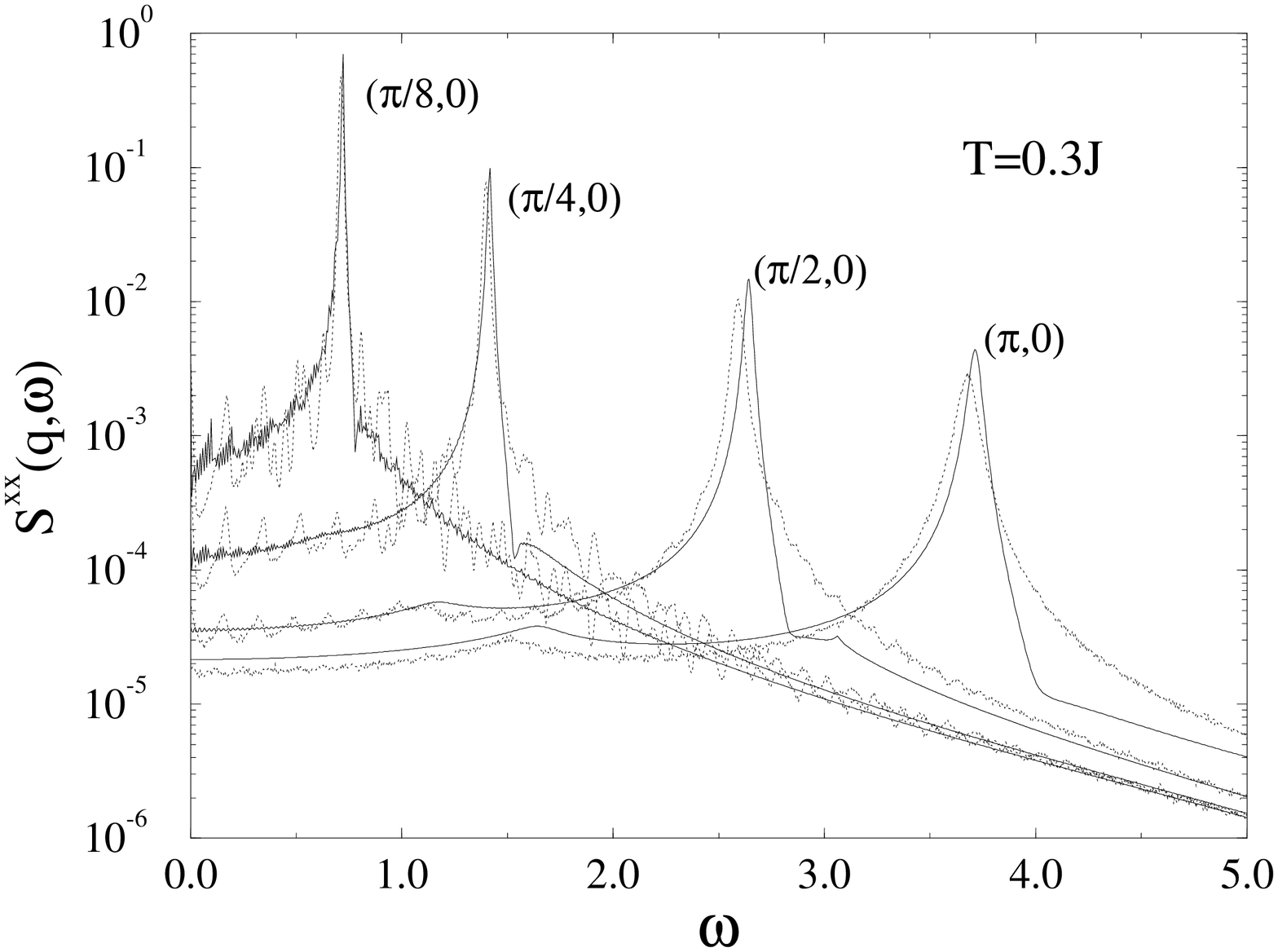,angle=-90.0,width=\pssize pt}
\hskip 1.6in
\psfig{figure=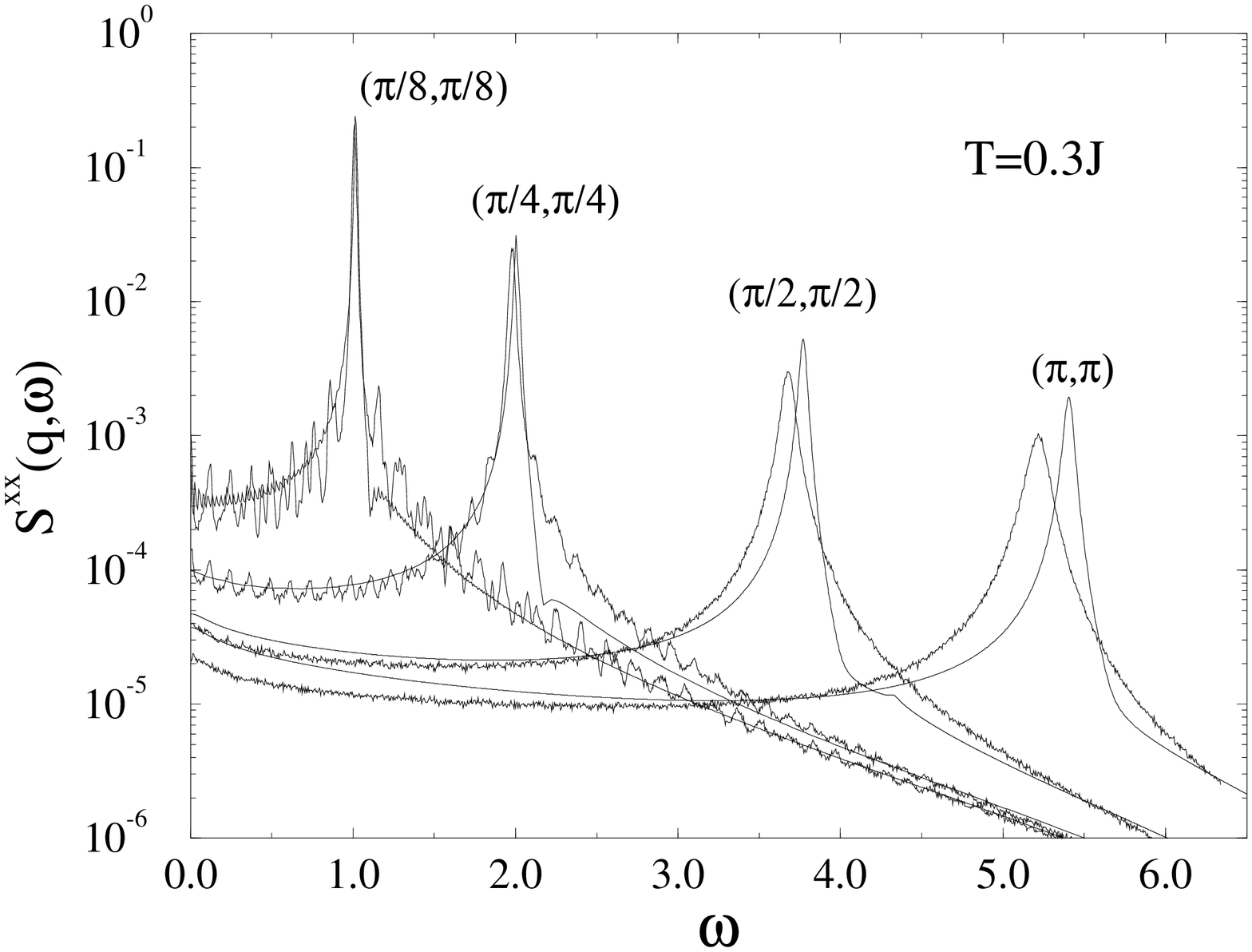,angle=-90.0,width=\pssize pt}
\caption{
\label{SxxT3-fig}
The $T=0.3 J$ dynamical structure function obtained from the memory
function calculations (solid curves, using $L=1024$--$2048$) 
compared with MC-SD data (dotted curves) for the $L=128$ system.
The peaks in the MC-SD data fall at slightly lower frequencies than
the memory predictions.
}
\end{figure}

\begin{figure}
\hskip 1.6in
\psfig{figure=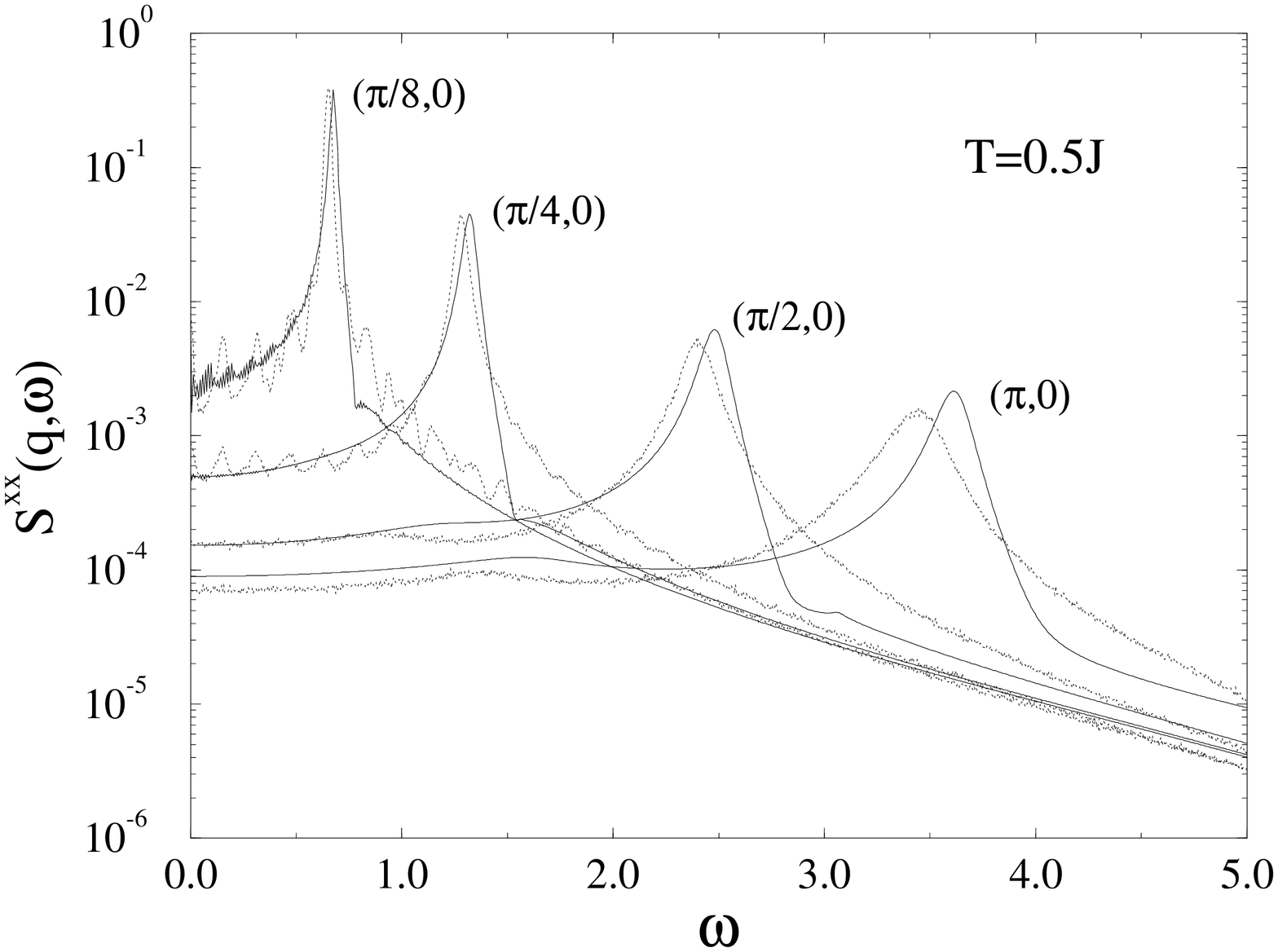,angle=-90.0,width=\pssize pt}
\hskip 1.6in
\psfig{figure=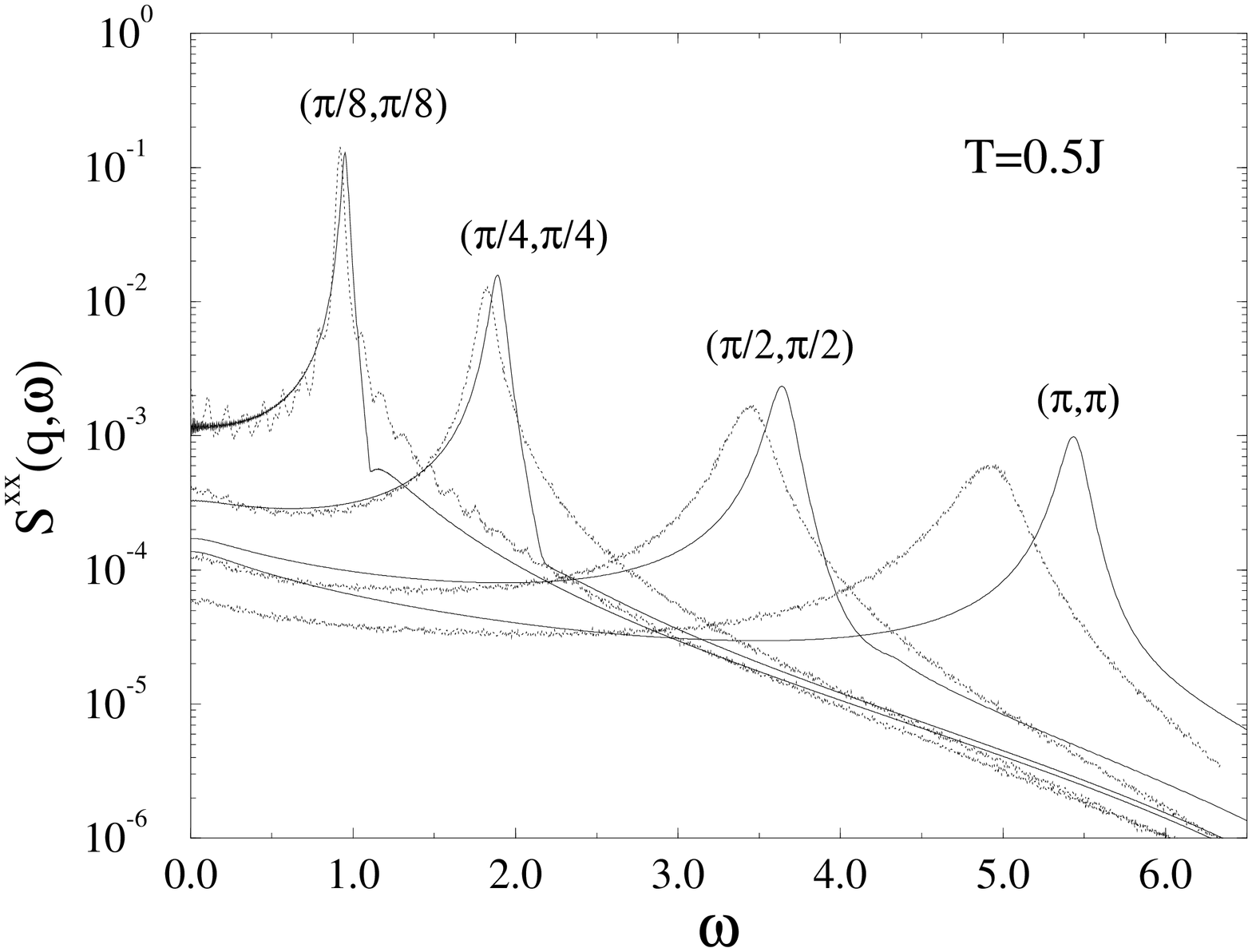,angle=-90.0,width=\pssize pt}
\caption{
\label{SxxT5-fig}
The $T=0.5 J$ dynamical structure function obtained from the memory
function calculations (solid curves, using $L=1024$--$2048$) 
compared with MC-SD data (dotted curves) for the $L=128$ system.
}
\end{figure}

In Figs.\ \ref{SxxT1-fig}--\ref{SxxT5-fig}, we display some further 
comparisons of the memory function calculations and MC-SD results at 
other values of $\bq$ for temperatures ranging from $T=0.1J$ to $T=0.5J$.
For the lower temperatures there is very good agreement between the
memory theory and MC-SD peak positions,  while the peak heights and
widths do not agree as well.
On the other hand, at higher temperatures the opposite is true:
the memory data show greater errors  in the peak positions,
especially at large $q$, but the peak heights and widths are 
predicted very well.
Probably these discrepancies are caused by a fairly large 
{\em underestimate} of the damping strength at low temperature, 
as well as errors in the estimate of the second moment at 
higher temperatures.  
However, in general, the tails of the spinwave peaks far from 
$\omega_{\bq}$ are described very well by the memory function
calculations.

\subsection{Dynamics: Small Systems and Finite Size Effects}
When we consider the results using relatively smaller values of
wavevector, $S^{xx}(\bq,\omega)$ shows not only a spinwave peak, 
but additional features (small peaks) which depend on the system size.
These features are obvious in the MC-SD $S(\bq,\omega)$ data 
shown above.
Similar size-dependent dynamical features have been noticed in earlier
MC-SD simulations of XY-symmetry\cite{GouveaWysin,Evertz,Landau99}
and easy-axis symmetry\cite{Gouvea+99} magnets,
but without a clear theoretical explanation.
We now consider a closer look at the details of these features, 
making a comparison with the memory function calculations
at smaller values of $L$.

\begin{figure}
\hskip 1.6in
\psfig{figure=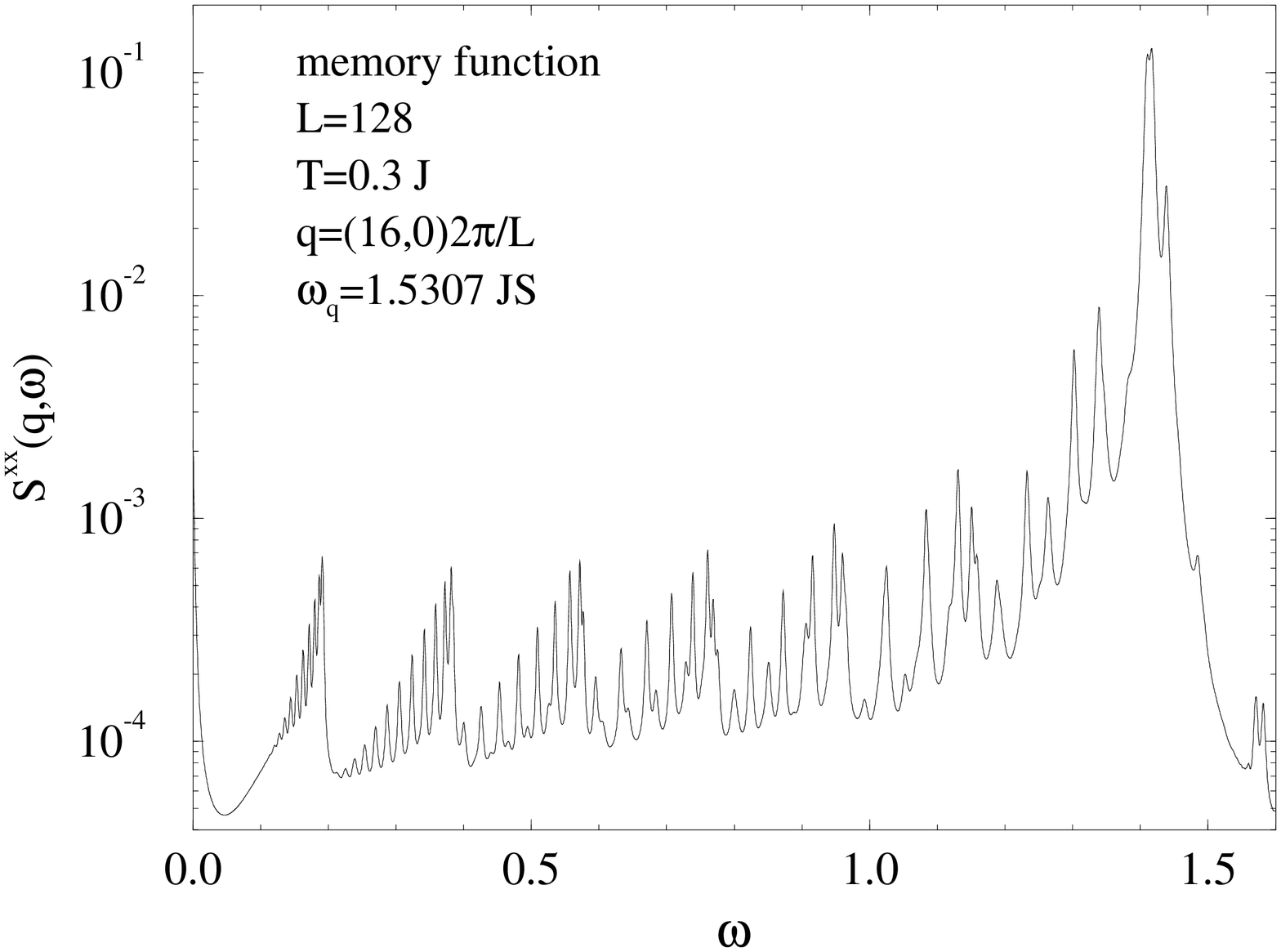,angle=-90.0,width=\pssize pt}
\hskip 1.6in
\psfig{figure=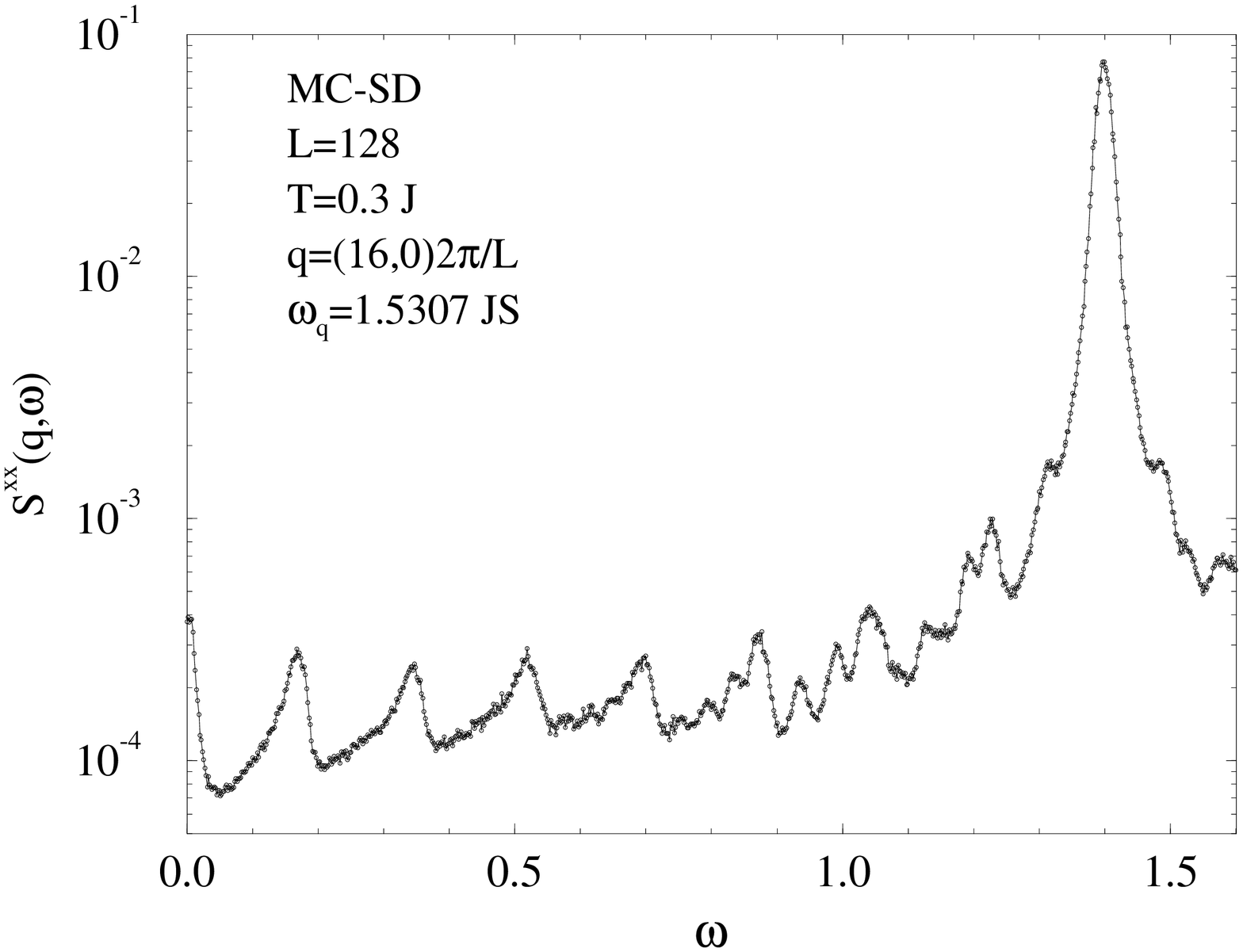,angle=-90.0,width=\pssize pt}
\caption{
\label{SxxT3q16}
Sequences of finite-size features in the dynamic structure function 
for an even integer wavevector, $\bq=(16,0)2\pi/L$, at $T=0.3 J$, from 
a) the memory function approach, and 
b) the Monte Carlo Spin-Dynamics.   
The frequency resolution of the SD data here and in
Figs.\ \protect\ref{SxxT3q9}, \protect\ref{SxxT1q5} is
$\Delta\omega = 0.0018262 JS$.
}
\end{figure}

\begin{figure}
\hskip 1.6in
\psfig{figure=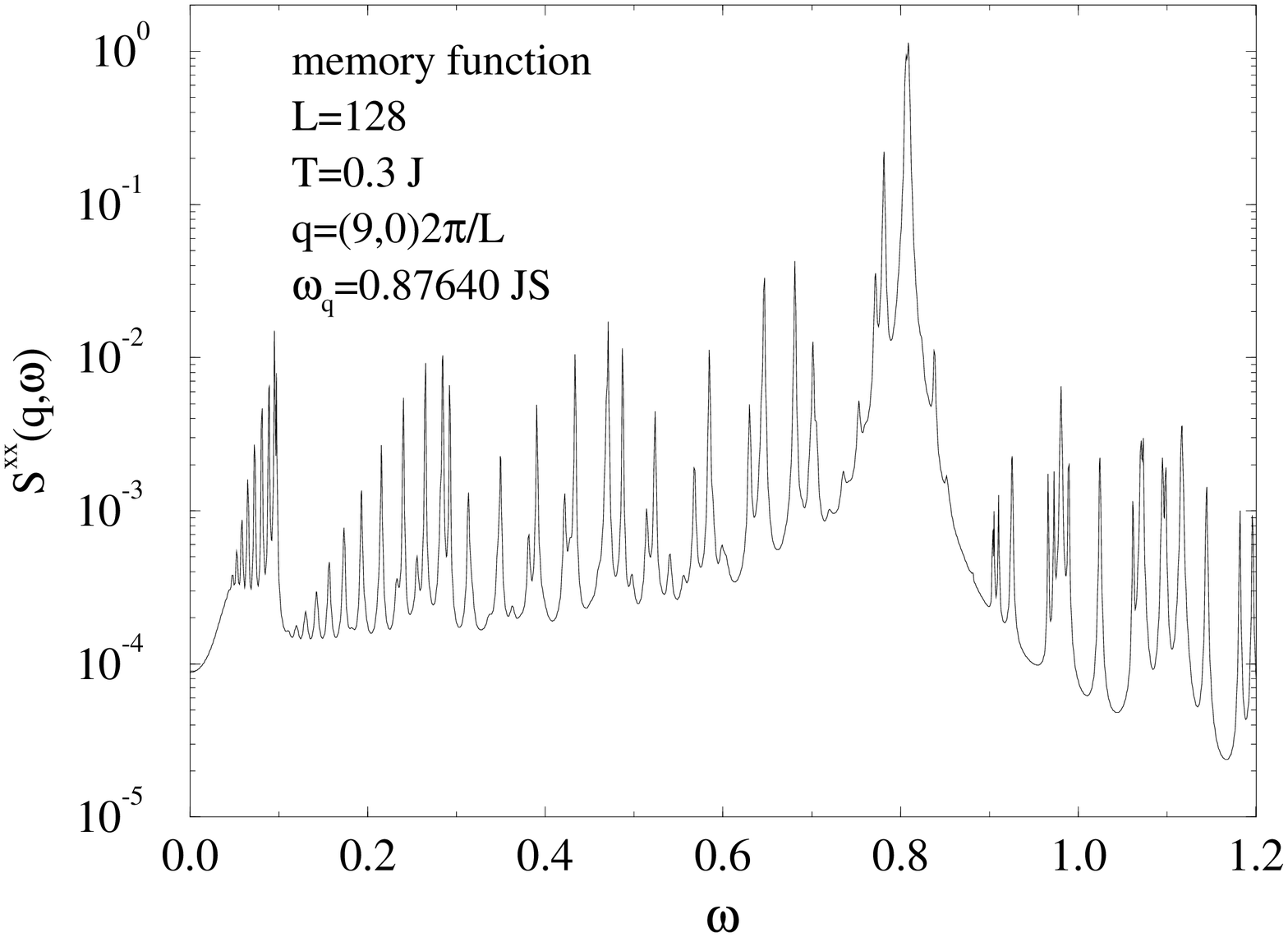,angle=-90.0,width=\pssize pt}
\hskip 1.6in
\psfig{figure=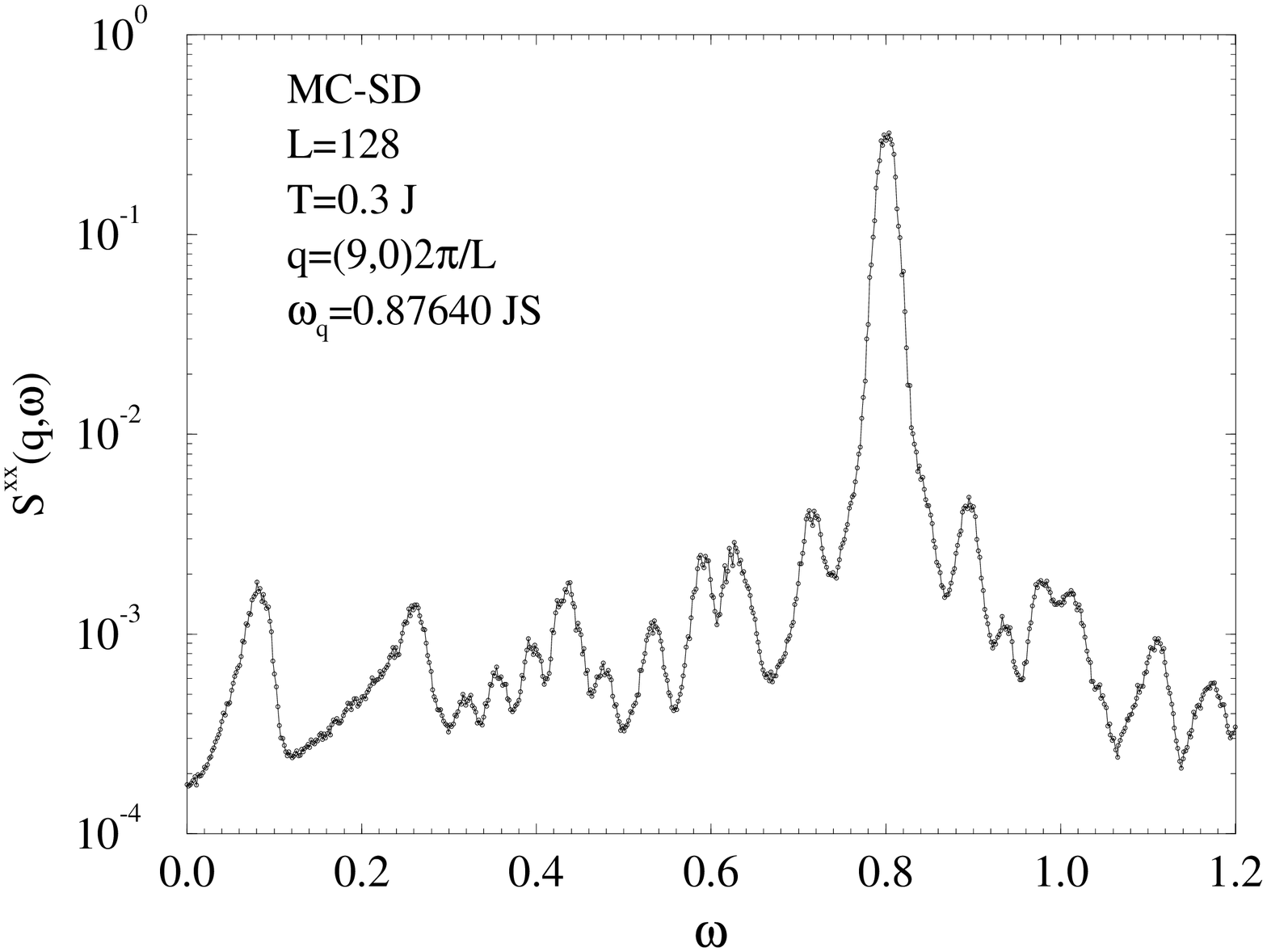,angle=-90.0,width=\pssize pt}
\caption{
\label{SxxT3q9}
Dynamic structure function for an odd integer wavevector,
$\bq=(9,0)2\pi/L$,  at $T=0.3 J$, from 
a) the memory function approach, and
b) the Monte Carlo Spin-Dynamics. 
}
\end{figure}

\begin{figure}
\hskip 1.6in
\psfig{figure=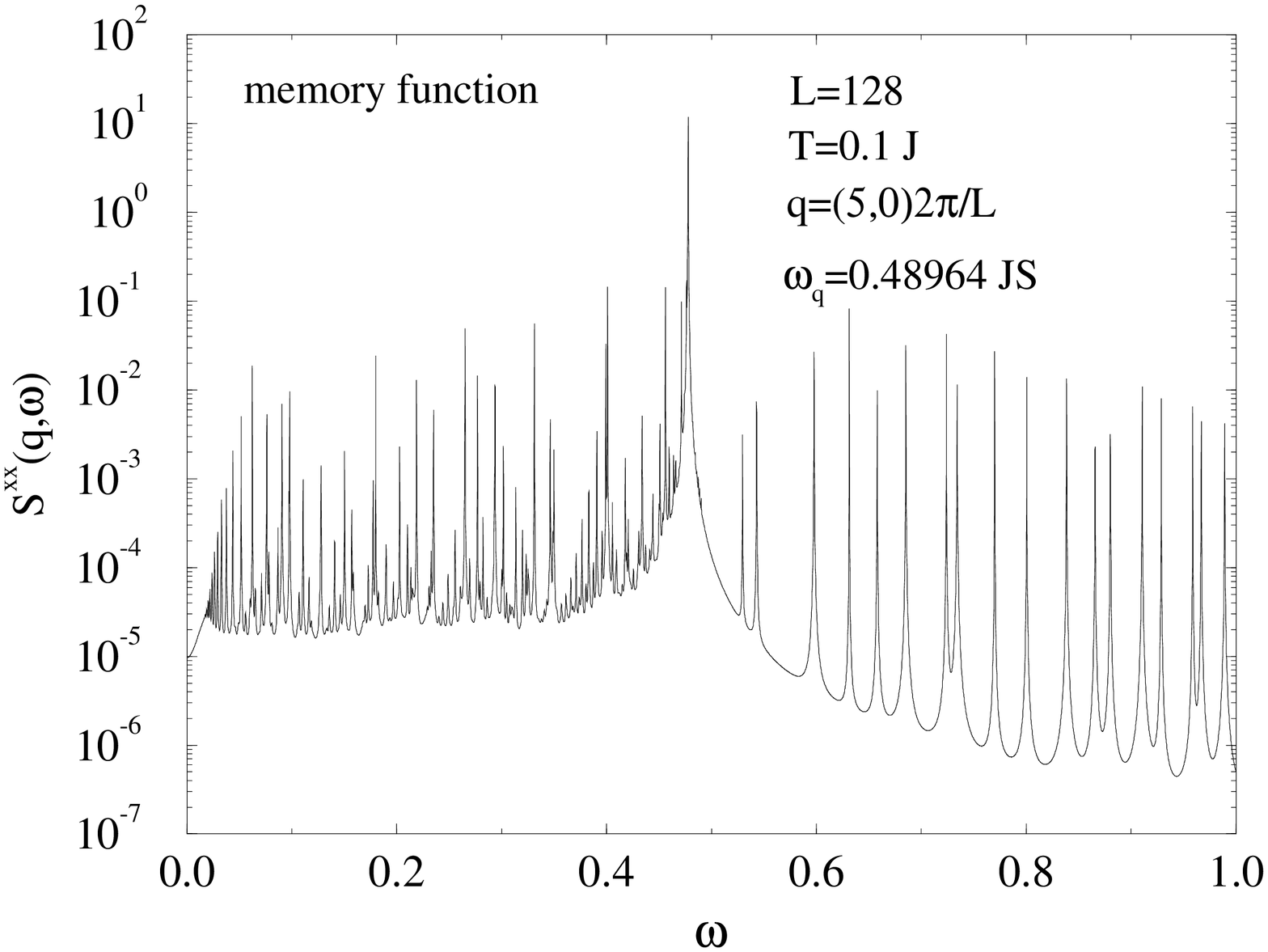,angle=-90.0,width=\pssize pt}
\hskip 1.6in
\psfig{figure=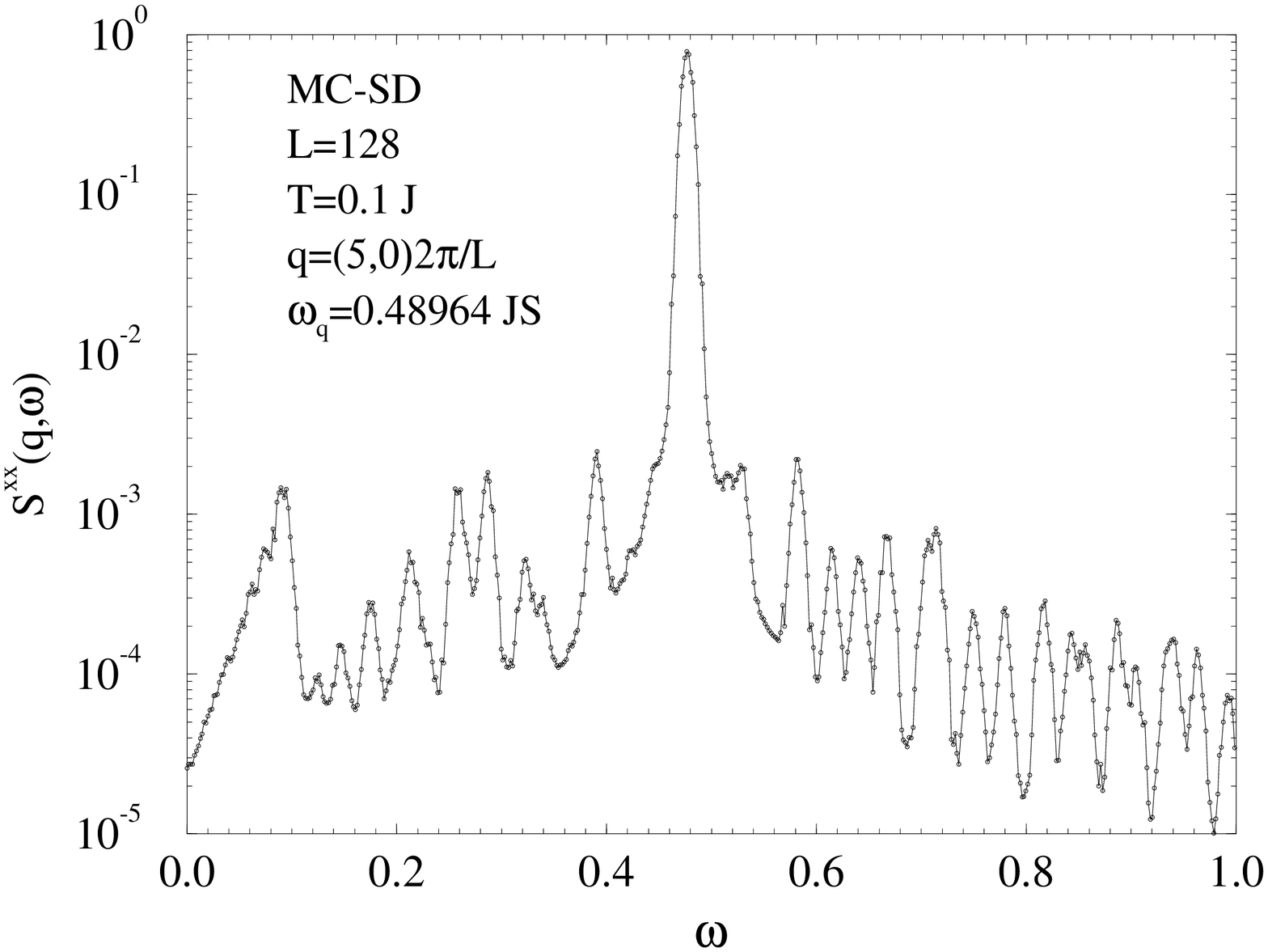,angle=-90.0,width=\pssize pt}
\caption{
\label{SxxT1q5}
Dynamic structure function for an odd integer wavevector,
$\bq=(5,0)2\pi/L$,  at $T=0.1 J$, from 
a) the memory function approach, and
b) the Monte Carlo Spin-Dynamics. 
}
\end{figure}

Some examples showing the resulting memory function data for 
$S^{xx}(\bq,\omega)$ for $L=128$ systems are given in
Fig.\ \ref{SxxT3q16}a [$\bq=(16,0)2\pi/L$, $T=0.3 J$],
Fig.\ \ref{SxxT3q9}a [$\bq=(9,0)2\pi/L$, $T=0.3 J$],
and Fig.\ \ref{SxxT1q5}a [$\bq=(5,0)2\pi/L$, $T=0.1 J$].
In fact, there appear various sequences of small peaks.
The number of separate sequences appears to depend on the
wavevector in integer units of $2\pi/L$.
Note that the logarithmic scale exaggerates the size of these
subpeaks, they are typically a few orders of magnitude smaller
than the spinwave peak.
For sufficiently low temperatures, these effects also appear
clearly in the MC-SD results: see Figs. \ref{SxxT3q16}b,
\ref{SxxT3q9}b and \ref{SxxT1q5}b, 
although the corresponding features there are 
smeared out compared to the memory function calculation.
The effects are also present at larger wavevectors, however, 
in that case there are many more of these finite size subpeaks,
which tend to merge together into a smoother background.

\begin{figure}
\hskip 1.6in
\psfig{figure=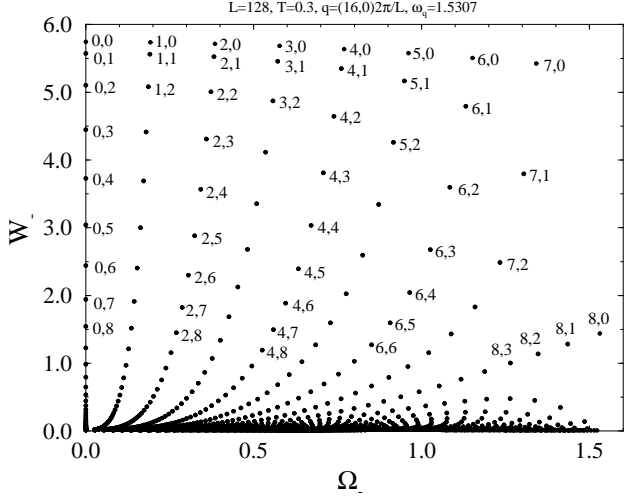,angle=-90.0,width=\pssize pt}
\caption{
\label{OmMinusT3q16}
Plot of difference process weights versus spinwave difference
frequency $\Omega_{-}=\omega_{\bk}-\omega_{\bq-\bk}$ for
the $L=128$ system, for an even integer wavevector 
$\bq=(16,0)(2\pi/L)$,  at $T=0.3 J$, showing sequences of 
difference processes with high weights, responsible for the 
sequences of finite size features in Fig.\ \protect\ref{SxxT3q16}. 
Numbers in parenthesis are wavevector values $\bk$ in units
of $2\pi/L$.
}
\end{figure}

These extra peaks can each be associated
with particular spinwave sum or difference process. 
To see that this is the case, consider first the difference
processes, whose frequencies $\Omega_{-}(\bq,\bp)$ are always less
than the spinwave frequency $\omega_{\bq}$.
A plot of the weights $W_{-}$ [Eq.\ (\ref{weights})] that determine 
the contributions to the memory function versus the corresponding 
frequencies $\Omega_{-}=\omega_{\bk}-\omega_{\bq-\bk}$, can
show how strongly each process [i.e., each $\bk$ as 
defined in (\ref{k-discrete})]
contributes to Im $\Sigma_{\bq}^{\perp}$
and hence to $S(\bq,\omega)$.
In Fig.\ \ref{OmMinusT3q16}, we made such a plot for an even
integer $\bq$, namely, $\bq=(16,0)(2\pi/L)$, for $L=128$ and $T=0.3J$.
Each dot there corresponds to one allowed wavevector $\bk$; some of 
the values of $\bk$ in units of $2\pi/L$ are indicated there. 
The sequence of $\bk$'s, (8,0), (8,1), (8,2), etc., leads to
multiple contributions at zero frequency, that is, a weak central peak,
which is a general result for any even integer $\bq$.
The sequence, (9,0), (9,1), (9,2), etc., produces a sequence
of small peaks in $S(\bq,\omega)$ out to $\omega\approx 0.2 JS$,
and, the next sequence, (10,0), (10,1), (10,2), etc.,
gives a sequence of peaks out to $\omega\approx 0.4 JS$,
and so on.
Each of these $\bk$-sequences can be identified with corresponding
peak-sequences in the memory function calculation of $S(\bq,\omega)$,
as seen in Fig.\ \ref{SxxT3q16}a. 
Indeed, for the wavevector $\bq=(16,0)(2\pi/L)$, there are $8+1$
of these sequences below the spinwave peak, including
the ones that contribute only at $\omega=0$. 
The higher sequences are more spread out, and merge with the
other ones, making a complicated fine structure in $S(\bq,\omega)$.
Similar features are seen in the MC-SD data, Fig.\ \ref{SxxT3q16}b, 
although it is clear that the individual peaks of each sequence
are smeared out, resulting in one wider feature from the entire sequence.
%
%

\begin{figure}
\hskip 1.6in
\psfig{figure=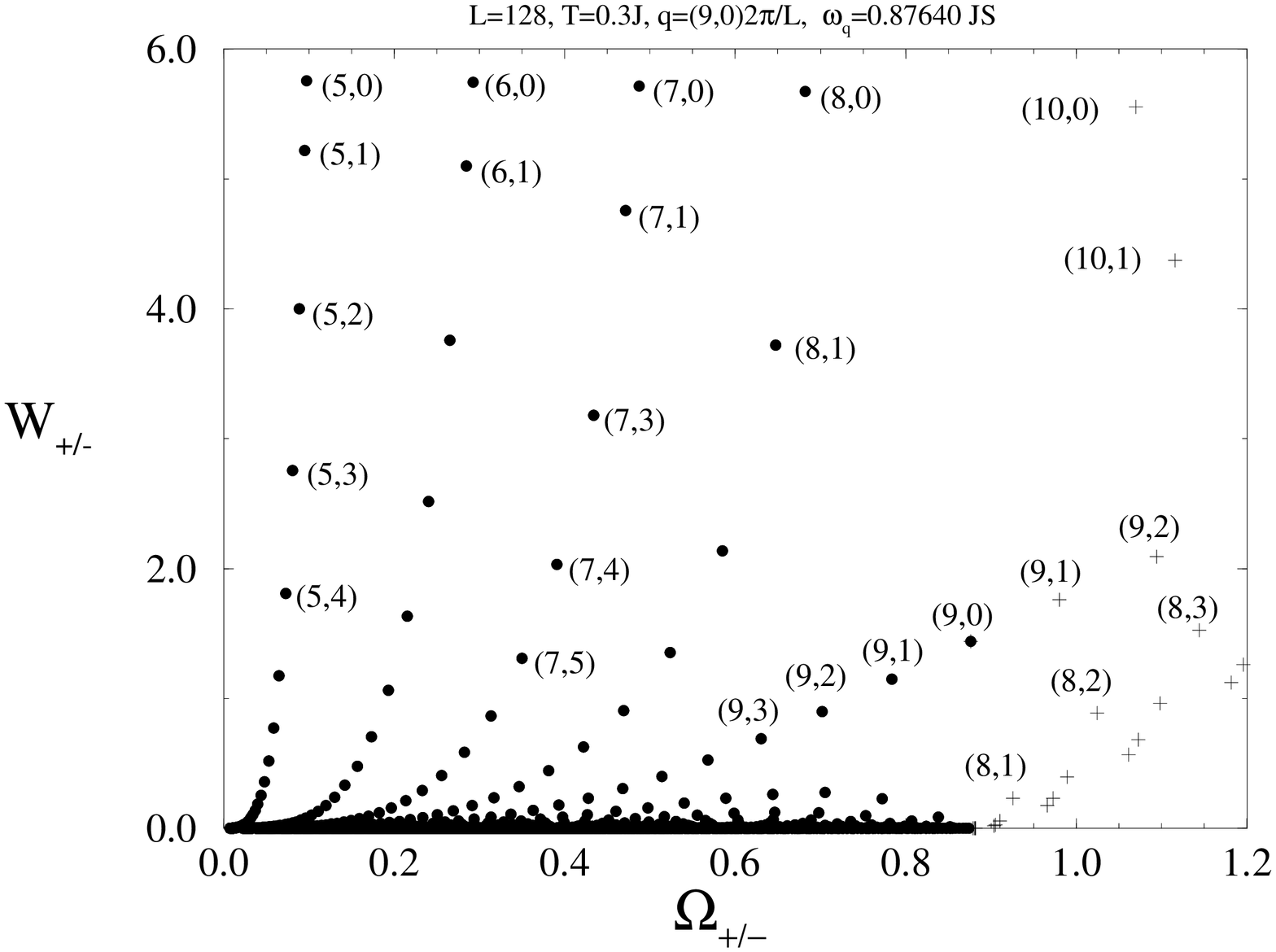,angle=-90.0,width=\pssize pt}
\caption{
\label{OmPMT3q9}
Sum (+ sigms)  and difference (solid circles) process weights versus 
frequencies $\Omega_{\pm}=\omega_{\bk}\pm\omega_{\bq-\bk}$ for
the $L=128$ system, for an odd integer wavevector 
$\bq=(9,0)(2\pi/L)$,  at $T=0.3 J$.  Compare the locations
of the features in Fig.\ \protect\ref{SxxT3q9}.
Numbers in parenthesis are wavevector values $\bk$ in units
of $2\pi/L$.
}
\end{figure}

Another example of these finite size features, for an odd integer
wavevector, $\bq=(9,0)(2\pi/L)$ is shown in Fig.\ \ref{OmPMT3q9},
again for $L=128$ and $T=0.3J$. 
For this smaller integer wavevector there are now only 5 
sequences of fine-structure difference process peaks.
It is significant that there is no such peak at zero frequency.
This is clearly caused by the absence of any allowed
spinwave difference process exactly at $\Omega_{-}=0$,
although there are many such processes with very weak
weights {\em close} to $\Omega_{-}=0$.
In general, for any odd integer wavevector, $S(\bq,\omega)$
will have a relative minimum at $\omega=0$, because
of the absence of allowed difference processes there.

In Fig.\ \ref{OmPMT3q9} we also show some of the weights
of the sum-processes, $\Omega_{+}=\omega_{\bk}+\omega_{\bq-\bk}$,
indicated by plus signs.
For either sum or difference processes, numbers in parenthesis
refer to the values of $\bk$ in units of $2\pi/L$.
The sum frequencies must fall above the spinwave frequency $\omega_{\bq}$.
Comparing Figures \ref{OmPMT3q9} and \ref{SxxT3q9}, we see that the
finite size features above $\omega_{\bq}$ in $S(\bq,\omega)$
can each be identified with a particular spinwave sum process.
The identification, however, is somewhat more complex than 
that for the difference processes,  because the different
sequences of processes overlap in frequency.
The match between the sum peak frequencies in the memory data and the 
MC-SD data is not as good as for the differences processes below $\omega_{\bq}$. 
We can note that at these higher frequencies, any relative errors
appear larger on an absolute scale, and, the spinwave sum/difference
frequencies do not include any exchange softening effects.

\begin{figure}
\hskip 1.6in
\psfig{figure=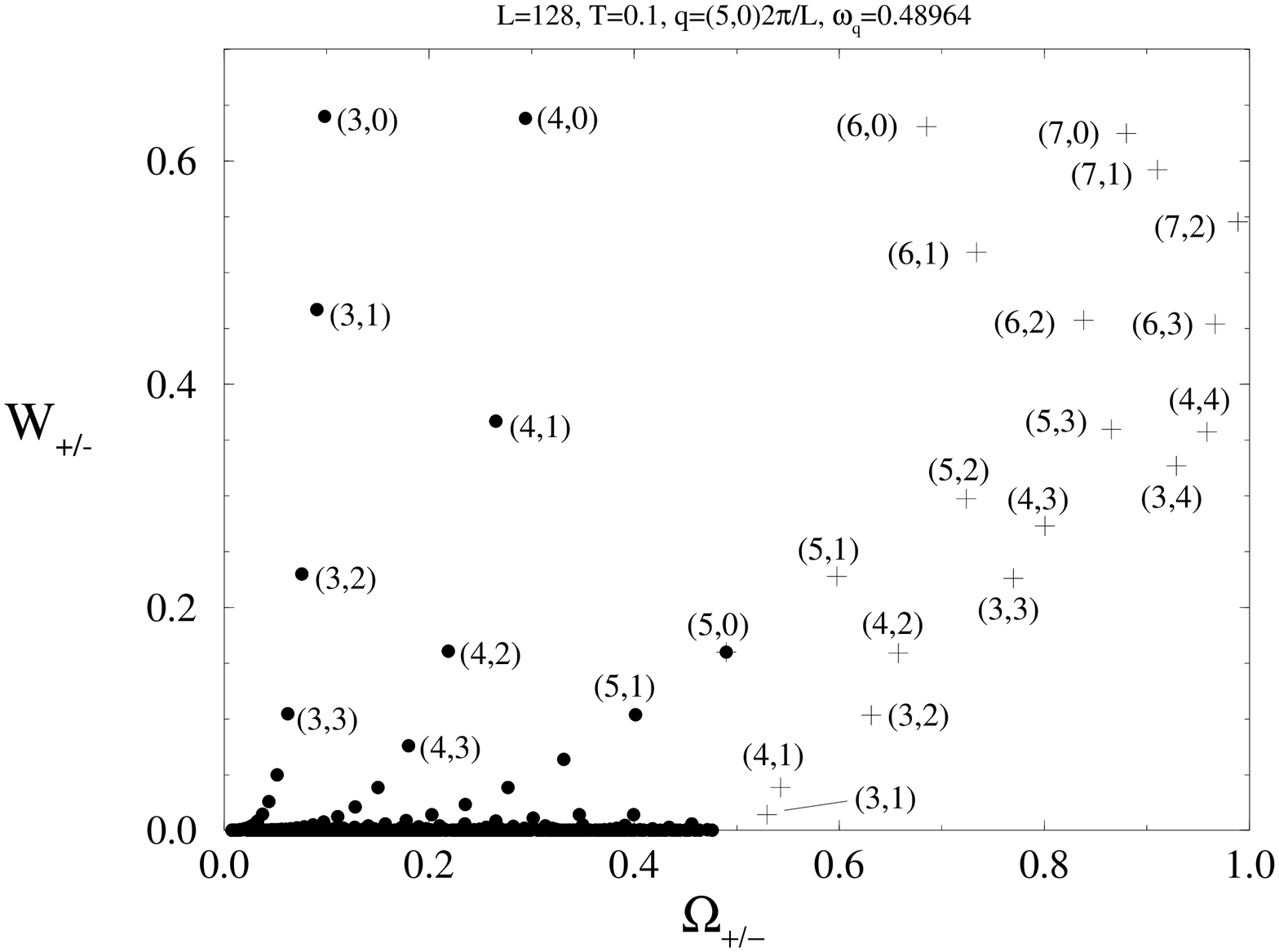,angle=-90.0,width=\pssize pt}
\caption{
\label{OmPMT1q5}
Sum (+ signs) and difference (solid circles) process weights versus
frequencies $\Omega_{\pm}=\omega_{\bk}\pm\omega_{\bq-\bk}$ for
the $L=128$ system, for an odd integer wavevector 
$\bq=(5,0)(2\pi/L)$,  at $T=0.1 J$.  Compare the locations
of the features in Fig.\ \protect\ref{SxxT1q5}.
Numbers in parenthesis are wavevector values $\bk$ in units
of $2\pi/L$.
}
\end{figure}

Finally we show another example, for $\bq=(5,0)(2\pi/L)$
at $T=0.1 J$, $L=128$ in Fig.\ \ref{OmPMT1q5}, where both
the difference and sum processes are shown.
In this case there are only 3 dominant sequences of difference
processes leading to finite size features.
%

\subsection{Higher Temperature}
Although the memory function formalism applied above, strictly speaking,
applies to low temperatures, it is interesting to consider what kind
of predictions it makes if applied at higher temperatures.
It is clear from the above comparisons between MC-SD data and
the memory function data that at low temperatures, where the
spinwave approximation should be valid, the damping of the
modes has been underestimated. 
This is seen especially in the $T=0.1J$ memory data where all the
finite size features, as well as the spinwave peak, are too
narrow.
Thus it is interesting to see whether the spinwave width and
shape are better described by this theory at higher temperature,
because the theoretical damping rates grow as $T^2$.
The other reason to look at higher temperature is to consider
how the background intensity in $S(\bq,\omega)$ for frequencies
$\omega<\omega_{\bq}$, depends on the temperature.
For one thing, the increase in temperature should spread all of 
the finite-size effect peaks and smooth the spinwave background,
but can be expected to raise its intensity as well.
For comparison, as $T$ crosses the vortex unbinding temperature
$T_{\rm BKT}$, previous MC-SD simulations\cite{MC-SD-Refs,Evertz,Landau99} 
show that this background grows with $T$, until it dominates over the 
spinwave peak.
The effect has been described by a theory employing a gas of freely
moving vortices,\cite{Mertens+89}  although simulations have given some
indication that vortex lifetimes\cite{lifetimes} are rather too short to 
take this theory literally, and, vortices tend to be pinned by the lattice
and also have complex motions not resembling free translation.
Thus it is interesting to compare the memory approach to the
MC-SD data at higher temperatures, where the latter will
include effects due to all possible types of excitations,
beyond linear spinwaves.

\begin{figure}
\hskip 1.6in
\psfig{figure=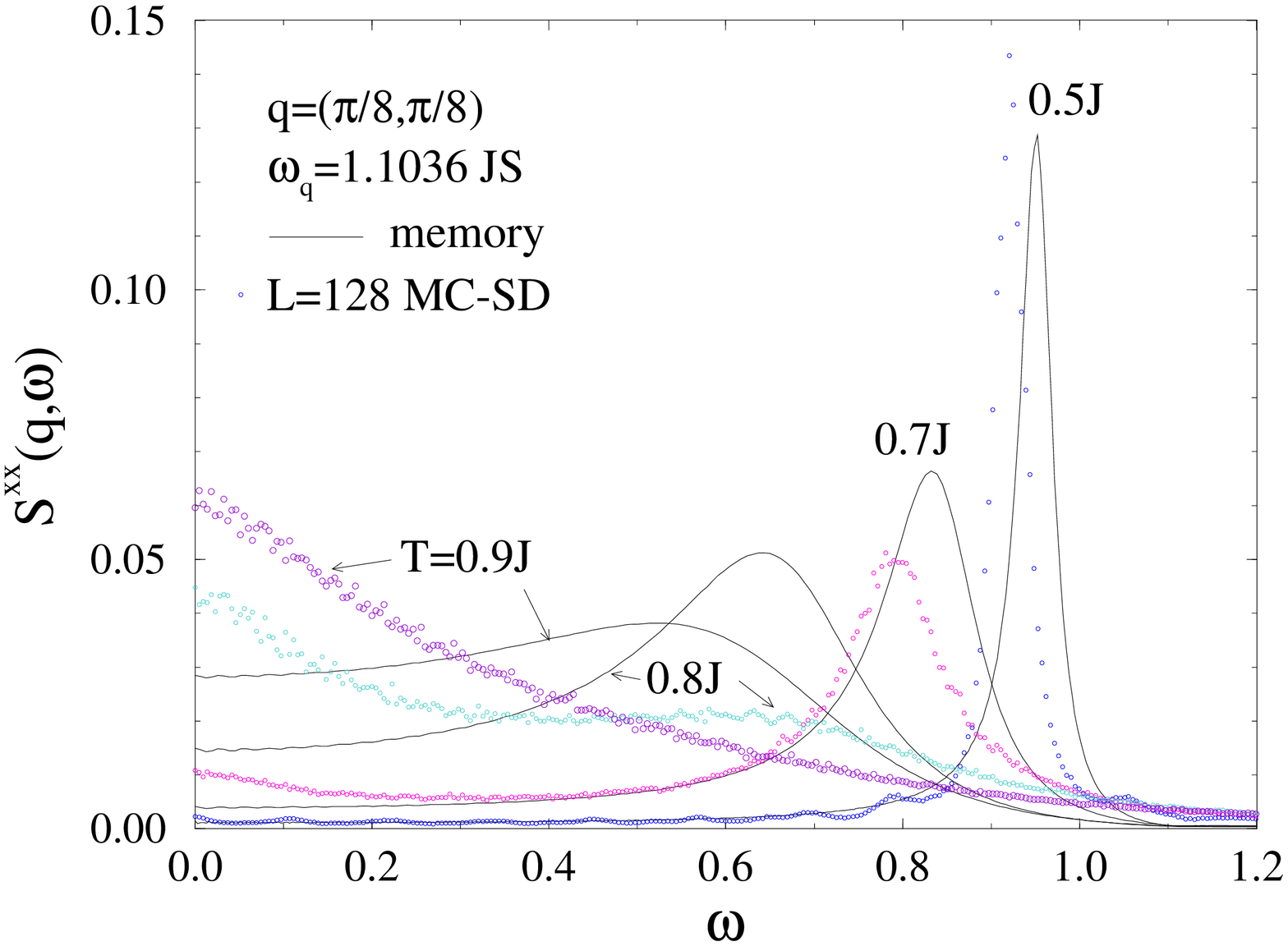,angle=-90.0,width=\pssize pt}
\hskip 1.6in
\psfig{figure=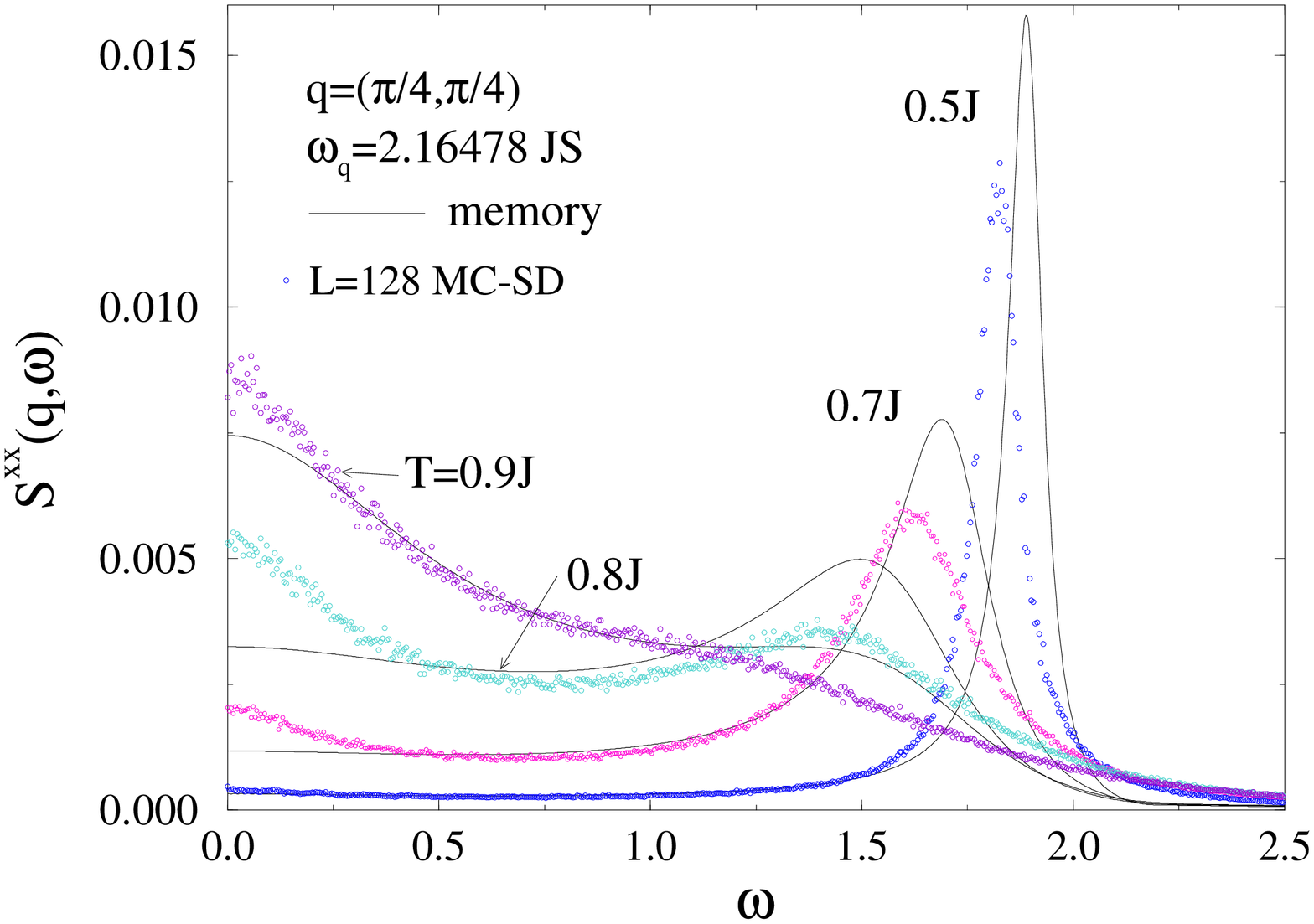,angle=-90.0,width=\pssize pt}
\caption{
\label{hi-T-Sxx}
Dynamic structure function at wavevectors $\bq=(\pi/8,\pi/8)$
and $\bq=(\pi/4,\pi/4)$, from the memory calculations (solid curves)
compared to Monte Carlo Spin-Dynamics (data points)
on $L=128$ systems, for a range of higher temperatures.
}
\end{figure}

For $\lambda=0$, the vortex unbinding temperature has been estimated to be 
$T_{\rm BKT}\approx 0.7 J$.
In Fig.\ \ref{hi-T-Sxx}, we show some comparisons of the 
continuum ($L \ge 1024$) memory  and MC-SD ($L=128$) data for 
$S(\bq,\omega)$, at temperatures $0.5J \ge T \le 0.9J$
for two particular values of $\bq$.  
As already seen,  the spinwave peak locations in the MC-SD data
here also fall lower than the memory function theory prediction.
Also, there is a significant growth of the low-frequency 
background ($\omega<\omega_{\bq}$) in the MC-SD data and in the memory 
function calculations.
For $T=0.9 J$ the low-frequency background already dominates
over the spinwave peak, if the latter can be seen at all.
The effect is stronger in the MC-SD data: there is a clear excess
of intensity around zero-frequency in $S(\bq,\omega)$ in the
MD-SD data compared to the memory function data.
This effect is absent at $T\le 0.5J$, however, where the
MC-SD and memory data coincide for low frequencies; in
this low temperature region the spinwave approximations used
in the memory approach are valid.
For temperatures near and above $T_{\rm BKT}$, this
extra central peak intensity, not described within the 
spinwave picture of the memory function approach, may
possibly be attributed to nonlinear excitations, including 
specifically, vortices.
This is consistent with the phenomenological vortex gas 
description of Mertens et al:\cite{Mertens+89}  the excess central 
peak component is stronger for lower wavevectors, because the 
vortex pairs being generated above $T_{\rm BKT}$ disorder the
spins on long length scales.   
%

\section{Discussion and Conclusions}
\label{Conclusions}
The presence of spinwave damping included into the memory
function approach has been suggested here as a way
in which finite amplitude spinwave peaks result in the 2D XY model,
canceling the divergences in previous theories.
A rough calculation of the damping strength has been made using
the Feynmann diagram technique applied to the quartic corrections
of the harmonic Hamiltonian.
The diagrams used involve only two scattering events; clearly
this gives an underestimate of the damping, and it is likely that
including the third order diagrams could make a substantial
modification to the results.
Generally the damping $\Gamma_{\bk}$ obtained increases away from 
the center of the Brilluoin zone, being close to zero near the 
center and being greatest along the (11) direction.
In this classical calculation, the damping varies as the square of
the temperature;  this results in extremely weak damping strength
($\Gamma_{\bk} \ll \omega_{\bk}$) for temperatures below $0.3J$.
As the temperature approaches and passes $T_{\rm BKT}$, the damping
becomes comparable to the spinwave frequency.
%

The comparison of the memory function data and MC-SD data 
for $S(\bq,\omega)$ is good in several aspects:  the theory
gives finite spinwave peaks, they are positioned correctly
for low temperatures, and the multiple finite-size-effect
peaks are reproduced as in the MC-SD data. 
The most significant problem in the memory function
approach is the extremely narrow peak widths--clearly
caused by an underestimate of the true damping strength,
especially at lower temperatures and wavevectors.
For very low temperatures in the present theory,
the longest wavelength spinwaves will not scatter even
once before crossing the entire system.
Of course, this is a quantum picture, applied to the classical system,
so perhaps a better approach in this low-T limit would be to 
consider some kind of adiabatic scattering in the
nonuniform potential due to the spinwave population.
This would essentially be the equivalent of the quantum 
calculation at very high order diagrams rather than at the
lowest order.
%

When the memory calculation is carried out for large systems, 
that is, in the continuum limit, the finite-size peaks merge into a
smooth background.
Indeed, in this limit, the spinwave peak itself has contributions on 
the high frequency side from spinwave sum processes, and on the low 
frequency side from spinwave difference processes.
The combination of all of these contributions, essentially an
infinite number of sum and difference processes, leads to an 
overall smooth and nondivergent spinwave peak.
Even in the limit of {\em zero damping}, the spinwave
peak are nondivergent, according to continuum calculations
[using Eqs.\ \ref{Sigma_int}] by Pereira et al.\cite{Pereira+00}
However, when damping is included, the heights of the spinwave peaks
are smaller, and we expect the results to be more realistic.

In the memory function theory, the spinwave peak height was seen to 
be proportional to $[{\rm Im}\Sigma(\omega_{\rm peak})]^{-1}$.
The function $\Sigma(\bq,\omega)$, to a leading approximation,
is proportional to the temperature $T$.
From Eq.\ (\ref{Sqw-memory}), in regions away from the spinwave
peak $S(\bq,\omega)$ should be proportional to $\Sigma(\bq,\omega)$.
Thus we see that the memory function theory demonstrates
how the background grows with temperature while the
spinwave peak relatively tends to fall.
For high enough temperature, the distribution of
spinwave difference processes near zero frequency dominates
the spectrum of $S(\bq,\omega)$, leading to a central
peak stronger than the spinwave peak.  
Clearly this means applying the theory well above the
limiting temperature where we expect spinwave approximations
to be valid.
However, this gives results very similar to those in the
MC-SD simulations, although the latter can be expected to
include nonlinear effects such as vortices.
In fact, we have seen a strong indication for nonlinear
effects (beyond spinwave difference processes)
above $T_{\rm BKT}$ as extra intensity in $S(\bq,\omega)$
near zero frequency in the MC-SD data.

In conclusion, application of the memory function formalism,
including a leading approximation to the mode damping, has
been shown to accomplish three things for the XY model:
i) produce nondivergent spinwave peaks;
ii) explain the finite size features in $S(\bq,\omega)$
both above and below $\omega_{\bq}$;
iii) explain the background in $S(\bq,\omega)$ below $\omega_{\bq}$
for the infinite system.
Note that strictly speaking, damping is necessary only for item ii),
because $S(\bq,\omega)$ cannot be calculated in the memory function
formalism for a finite system without damping.
The continuum memory function approach gives finite size spinwave
peaks and a low-frequency background even in the absence of damping.
The theory also suggests that a sizable contribution
to the growth of the central peak for $T$ above $T_{\rm BKT}$
may be due to spinwave difference processes.
The excess central peak intensity in the MC-SD data, when
compared to the memory function data,  must be due to 
nonlinear excitations, and gives evidence for dynamical effects 
due to vortices.
%

\medskip
{\sl Acknowledgments.}---GMW thanks Universidade Federal de Minas 
Gerais for their hospitality and support on this project. The authors
thank the support by NSF-INT-9502781,  NSF CDA-9724289, FAPEMIG, 
and CNPq.



\vskip 2.0in

\begin{references}
\bibitem{Gilat2d3d} In two-dimensions, Eq.\ (\protect\ref{Dcell}) becomes
\begin{equation}
D_{\rm cell}(\omega) =  \frac{f(x_c)}{\vert \nabla g(x_c) \vert}
\frac{1}{2 \vert l_1 l_2 \vert} \sum_{i=1}^{4} (-1)^{m_i}
\vert w + w_i \vert,
\end{equation}
%
and in three dimensions,
\begin{equation}
D_{\rm cell}(\omega) =  \frac{f(x_c)}{\vert \nabla g(x_c) \vert}
\frac{1}{4 \vert l_1 l_2 l_3 \vert} \sum_{i=1}^{8} (-1)^{m_i}
\vert w + w_i \vert (w + w_i).
\end{equation}
%
For arbitrary dimension $d$, the generalization  is
\begin{equation}
D_{\rm cell}(\omega) =  \frac{f(x_c)}{\vert \nabla g(x_c) \vert}
\frac{1}{2 (d-1)! \vert l_1 l_2 ... \vert} \sum_{i=1}^{2^d} (-1)^{m_i}
\frac{(w + w_i)^d} {\vert w + w_i \vert} 
\end{equation}
\bibitem{FTNote}
The usual conversion from discrete to continuum Fourier weights is to
divide by the volume $(2\pi/L)^{2}$ per allowed point in $\bq$-space:
%
\begin{equation}
F(\bq) = \left( \frac{L}{2\pi} \right)^{2} F_{\bq}
\end{equation}
\end{references}
\end{document}